\documentclass[lettersize,journal]{IEEEtran}
\usepackage{amsmath,amsfonts}
\usepackage{cite}
\usepackage{array}
\usepackage{textcomp}
\usepackage{stfloats}
\usepackage{url}
\usepackage{verbatim}
\usepackage{graphicx}  
\usepackage{subcaption}
\usepackage{textcomp}
\usepackage{xcolor}
\usepackage{multirow}
\usepackage{eurosym}
\usepackage[linesnumbered,ruled]{algorithm2e}
\usepackage{enumitem}
\usepackage{tablefootnote}
\usepackage{nomencl}
\makenomenclature
\IEEEoverridecommandlockouts                              
\allowdisplaybreaks

\newtheorem{thm}{Theorem}
\newtheorem{remark}[thm]{Remark}

\usepackage{etoolbox}
\renewcommand\nomgroup[1]{%
  \item[\bfseries
  \ifstrequal{#1}{P}{\textit{C. Variables}}{%
  \ifstrequal{#1}{N}{\textit{A. Sets}}{%
  \ifstrequal{#1}{O}{\textit{B. Parameters}}{}}}%
]}

\begin{document}
	
	\title{Distributed Coordination of Multi-Microgrids in Active Distribution Networks for Provisioning Ancillary Services} 
 
    \author{Arghya Mallick,~\IEEEmembership{Student Member,~IEEE,} Abhishek Mishra,~\IEEEmembership{Student Member,~IEEE,} Ashish R. Hota,~\IEEEmembership{Senior Member,~IEEE,} Prabodh Bajpai,~\IEEEmembership{Senior Member,~IEEE} 

    \thanks{This work is supported by Science and Engineering Research Board (SERB), Government of India, through IMPRINT-IIC scheme (Ref. No. IMP/2019/000451/EN), and SERB sponsored Center of Excellence on Energy Aware Urban Infrastructure, IIT Kharagpur (Ref. No. IPA/2021/000081). Abhishek Mishra is gratefully supported by Prime Minister's Research Fellowship. (\textit{Corresponding author: Arghya Mallick.})
    
    Arghya Mallick is with the Delft Centre for Systems and Control, TU Delft, Netherlands. Abhishek Mishra and Ashish R. Hota are associated with the Department of Electrical Engineering, Indian Institute of Technology (IIT) Kharagpur, West Bengal, India. Prabodh Bajpai is with Sustainable Energy Engineering Department of IIT Kanpur, India. Emails (\textit{respectively}): A.Mallick{@}tudelft.nl, abhishekmishra.ee21{@}kgpian.iitkgp.ac.in, ahota{@}ee.iitkgp.ac.in, and pbajpai{@}iitk.ac.in.}} 


\maketitle

\begin{abstract}
With the phenomenal growth in renewable energy generation, the conventional synchronous generator based power plants are gradually getting replaced by renewable energy sources (RESs) based microgrids. Such transition gives rise to the challenges of procuring various ancillary services from microgrids. We propose a distributed optimization framework that coordinates multiple microgrids in an Active Distribution Network (ADN) for provisioning passive voltage support based ancillary services while satisfying operational constraints. Specifically, we exploit the reactive power support capability of the inverters and the flexibility offered by storage systems available with microgrids for provisioning ancillary service support to the transmission grid. We develop novel mixed integer inequalities to represent the set of feasible active and reactive power exchange with the transmission grid that ensure passive voltage support. The proposed alternating direction method of multipliers (ADMM) based algorithm is fully distributed, and does not require the presence of a centralized entity to achieve coordination among the microgrids. We present detailed numerical results on the IEEE 33-bus distribution test system to demonstrate the effectiveness of the proposed approach and examine the scalability and convergence behavior of the distributed algorithm for different choice of hyper-parameters and network sizes.
\end{abstract}
	

\section{Introduction}
	
\IEEEPARstart{A}{ncillary} services (AS) such as frequency control, voltage control, and ramping support play an essential role in reliable and resilient power systems operation \cite{kumar2021ancillary,gautam2020resilience, quan2022distributed, sanandaji2015ramping,al2023dynamic}. Due to major shift towards RESs integrated power system, the synchronous generator based power plants are gradually phasing out resulting in diminished source of reactive power in the transmission network, which is poised to disturb voltage regulation of the system in long run \cite{karbouj2018voltage}. In view of voltage control based AS, transmission system operators (TSOs) have invested considerable capital in voltage regulating devices such as capacitor banks, static VAR compensators, and tap changing transformers. Although these devices perform effectively, they incur high capital cost which is uneconomical \cite{rousis2020provision}. Consequently, TSOs are increasingly seeking to tap into flexibility resources available at active distribution networks (ADNs) that include distributed energy resources (DERs), storage units, microgrids and smart buildings for procuring AS. However, extracting such flexibility from ADNs is quite challenging as it requires suitable coordination and optimization of these multitude of entities subject to stringent operational constraints. 

In connection with this, authors in \cite{stavros_as} have proposed a centralized optimal power flow problem to optimize distribution network operation and also provide voltage support to the transmission network considering operational constraints of the ADN and MGs. Similarly, a recent work \cite{florian_as} leverages the flexibility offered by smart buildings (with its local generation and storage units) present in the low voltage network to provide voltage support to the medium voltage network. The authors consider a hierarchical optimization approach where smart buildings solve their local energy management problems while responding to suitable signals sent by the ADN. Provisioning voltage ancillary services to the transmission system by utilizing static compensators for static and dynamic reactive power support was studied via simulations in \cite{rousis2020provision}. Furthermore, provisioning of reactive power support based AS is demonstrated in \cite{harighi2023provision} while authors in \cite{dutta2022coordinated} have developed a scheduling scheme for Volt-VAR devices to control bus voltage magnitude in ADN. An ESSs planning algorithm was proposed in \cite{abdeltawab2021energy} for providing localized reactive power support, peak shaving, and energy arbitrage related ASs. Authors in \cite{guo2018mpc} have introduced a coordinated voltage control scheme for ADNs to regulate bus voltages.

These above works are based on centralized optimization approaches, and have demonstrated how to leverage the flexibility of DERs in an ADN for provisioning of ASs. However, as the number of microgrids and smart buildings grow, such centralized techniques would not remain practical and scalable. Specifically, it is neither desirable nor feasible for a single entity to keep track of local generation, demand, state of charge of storage units, load curtailment preferences and other operational constraints of all constituent smart buildings and microgrids in an ADN to jointly schedule their operation. Increasing concern regarding user privacy is another aspect that may hinder deployment of such centralized schemes. 
	
In contrast with centralized optimization, in distributed optimization schemes the overall optimization problem is decomposed into smaller problems, each solved by an independent entity with limited peer-to-peer communication among such agents. In such distributed schemes, autonomy of the subsystems are preserved and their local information is not shared with other agents. In addition, such schemes are inherently scalable and fault-tolerant. Due to the above advantages, distributed optimization algorithms have been developed for several problems that arise in power systems; see \cite{molzahn2017survey} for an extensive review. More specifically, algorithms based on alternating direction method of multipliers (ADMM) \cite{boyd_admm, makhdoumi2017convergence} have received a lot of attention by the researchers. ADMM based distributed OPF was investigated in \cite{erseghe2014distributed, peng2016distributed}. A distributed energy management framework employing ADMM was developed in \cite{9393595} to obtain optimal scheduling of the energy resources in an ADN.  Authors in \cite{babagheibi2022distribution} have used ADMM while solving congestion management problem in ADN with multi-microgrids. 
		
However, to the best of our knowledge, distributed optimization techniques have not been studied in the context of provisioning ASs by ADNs except in a very recently published work \cite{quan2022distributed}. Authors in \cite{quan2022distributed} have used a distributed particle swarm optimization (PSO) based coordination framework which provides no guarantee of the optimality of solutions. In addition, their study does not provide any supporting results showing the error between centralized and distributed solutions which is very crucial to adjudge the quality of the distributed algorithm. In this paper, we address the above research gap and propose a distributed optimization approach for coordination of multiple microgrids in an ADN for efficient operation and provisioning of ancillary services. Our contributions are summarized below.
\begin{enumerate}
\item We formulate a multi-stage optimization problem where an ADN provides voltage support based AS while meeting operational constraints including line flow limits, voltage magnitude limits, microgrid inverter limits and constraints on battery energy storage systems (BESSs) (Section II). Furthermore, we present a set of mixed integer linear inequalities to capture the set of feasible active and reactive power exchange between the ADN and the TN that ensure penalty free operation under the passive voltage support scheme proposed by the Belgian TSO \cite{as_belgian}. Our proposed formulation corrects certain inaccuracies present in a similar formulation proposed in \cite{florian_as} where the set of feasible active and reactive power exchange was incorrectly modeled which would potentially lead to higher penalty and increased load curtailment in the network.   
\item An ADMM based distributed optimization framework is developed in order to solve the above optimization problem in a fully distributed manner following recent works \cite{dist_admm_1,makhdoumi2017convergence,admm_hota} (Section III). Under the proposed scheme, each microgrid determines its BESS charging schedule and load curtailment while the ADN is responsible for constraints pertaining to power flow and AS support. The proposed approach does not require the presence of a centralized entity (unlike past works such as \cite{liu2018novel}), and does not require exchange of dual variables among the agents (unlike \cite{peng2016distributed}). In particular, agents are only required to transmit information about active and reactive power exchange between MGs and ADN which are then used to achieve consensus.\footnote[1]{An added advantage of our approach is that unlike dual variables that correspond to shadow prices, active and reactive power are physical quantities, and hence less prone to manipulation, which is a growing concern in distributed optimization schemes \cite{sundaram2018distributed,tanaka2015faithful}.}   
\end{enumerate}
 
Detailed numerical results, obtained on the IEEE 33-bus distribution test system (Section IV), illustrate
\begin{itemize}
\item how the proposed scheme leverages inverter reactive power control capabilities (managed locally by MGs) for provisioning ASs (power exchange with TN managed by ADN), and comparison with the formulation developed in \cite{florian_as}, 
\item the effect of ADMM hyper-parameters on the convergence rate, and error between centralized and distributed solutions, and
\item the scalability of the proposed approach with an increasing number of microgrids and larger test networks  (with $69$, $136$, and $906$ buses). 
\end{itemize} 
We conclude with a discussion on promising directions for future research in Section V.

\section{Problem Formulation}
In this section, we develop a multi-stage optimization problem for optimal operation of an ADN that provides passive voltage support to the TN. Let the number of buses of the ADN be $N_b$, total number of distribution lines be $N_l$, and the number of microgrids connected to the ADN be $N_m$. Each microgrid consists of a Battery Energy Storage Systems (BESS), a PV generation unit, and local load with possibility of load curtailment. Thus, each microgrid could well represent a smart sustainable building. The prediction horizon for the problem is denoted by $N_p$. We define the set $[M] :=\{1,2, \dots M\}$ for brevity of notation. The decision vector at time $k \in [N_p] $ is defined as 
\begin{align*}
		\mathbf{x}_k= & [P_k, Q_k, P^{inj}_k, Q^{inj}_k, V^{sq}_k, I^{sq}_k, P^{ex}_k, Q^{ex}_k, P^{bat}_k, P^{curt}_k,\\
		& Q^{inv}_k, E_{k+1}, C^{tn}_k]^{\top},     
\end{align*}
where $P_k \in \mathbb{R}^{N_l}$ and $Q_k \in \mathbb{R}^{N_l}$ represent the active and reactive power flows in the lines of the ADN, $P^{inj}_k \in \mathbb{R}^{N_b}$ and $Q^{inj}_k \in \mathbb{R}^{N_b}$ denote the active and reactive power injections in the buses of the ADN, $V^{sq}_k \in  \mathbb{R}^{N_b} $ denotes the square of the magnitude of bus voltages, $I^{sq}_k \in  \mathbb{R}^{N_l}$ is the square of the magnitude of line currents, $P^{ex}_k \in \mathbb{R}$ and $Q^{ex}_k \in \mathbb{R}$ denote the active and reactive power exchange with the TN at the substation node (positive value implies import of power from TN and vice versa), $P^{curt}_k\in \mathbb{R}^{N_b}  $ refers to the load curtailment in the buses, $P^{bat}_k \in \mathbb{R}^{N_m}$ refers to the power supplied by the BESSs in the microgrids, $Q^{inv}_k\in \mathbb{R}^{N_m}$ is the reactive power generation / absorption by the inverters located in microgrids, $E_{k+1} \in \mathbb{R}^{N_m}$ refers to the available energy of BESS in MGs at time $k+1$ after $P^{bat}_k$ is withdrawn. Note that $E_1$ represents the energy stored at the current interval and is typically known from the state of charge provided by the battery management system (and hence not a decision variable for the proposed optimization problem). Finally, $C^{tn}_k$ is the cost of violating passive voltage constraint which will be subsequently defined. We now introduce the operational constraints acting on these decision variables.  Important notations are summarized in the nomenclature section.

\begin{figure}
    \centering
    \includegraphics[width=0.7\linewidth]{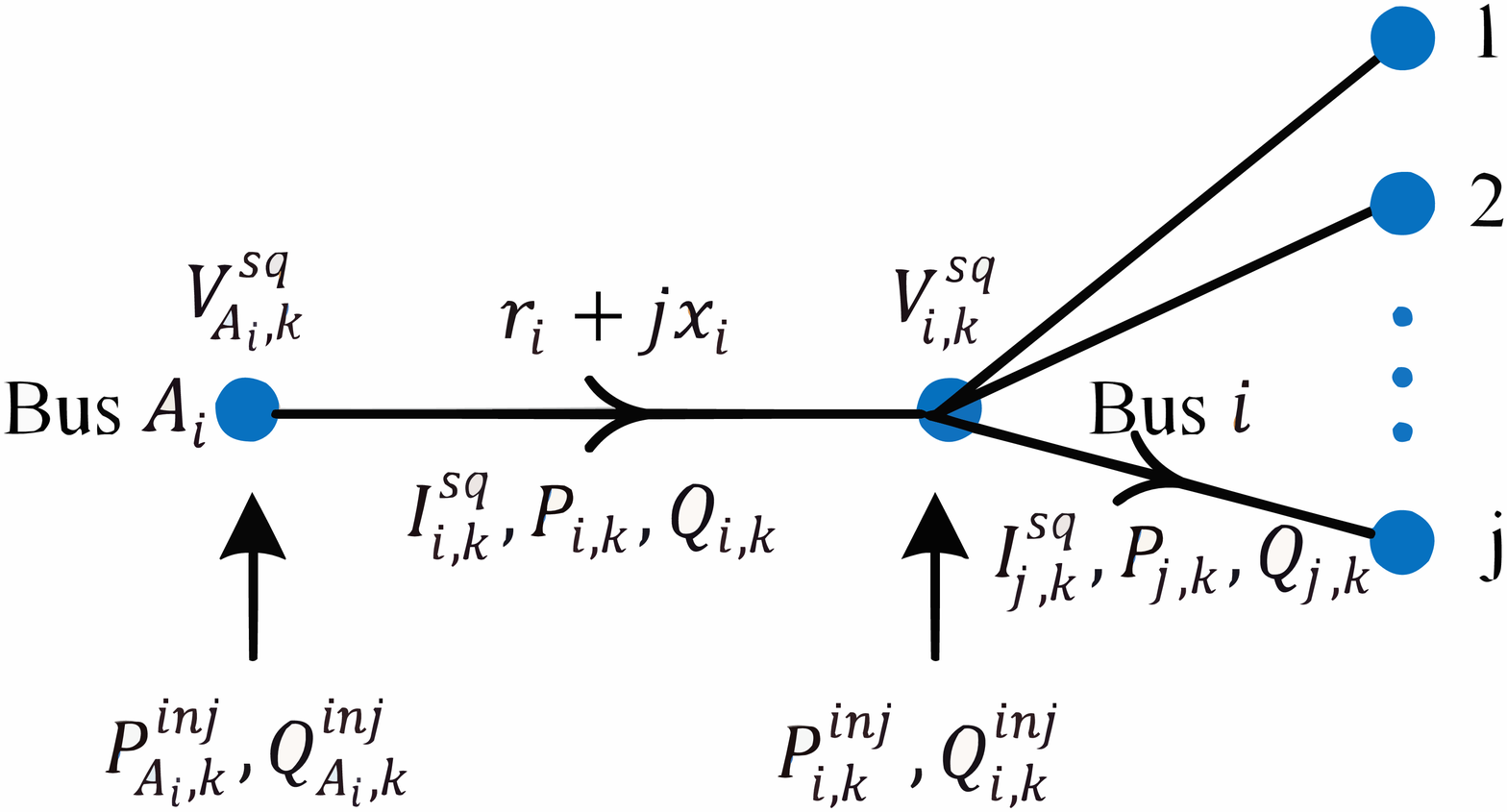}
    \caption{ Notations for branch flow model}
    \label{branchflow_Fig}
\end{figure}

\subsection{Power Flow Constraints}
We model the radial ADN using a directed tree graph $\mathcal{S} := (\mathcal{T,U}) $ where $\mathcal{T}$ represents the set of buses and $\mathcal{U}$ is the set of distribution lines. Each bus (or node) $i$ in the graph has an unique ancestor $A_i$ (except the substation node that connects the ADN with the TN), and a set of children nodes denoted by $\mathcal{C}_i$. The convention of graph orientation is such that every line points away from the substation or root node. In Fig. \ref{branchflow_Fig}, the orientation of the graph is shown where the direction of power flow in line $i$ is assumed to be from the ancestor $A_i$ to the node $i$. Other quantities such as nodal injection, line current and bus voltage are also shown in the figure. 

We have adopted the relaxed branch flow model (also known as \textit{Distflow} model) to represent power flow in the radial ADN \cite{farivar_distflow,peng_distflow}. According to \cite{peng_main}, branch flow model is more numerically stable than bus injection model. The branch flow equations can be written as, 
\begin{subequations}
		\begin{align}
			& P_{i,k} = \: \sum_{j\in  \mathcal{C}_i} \: \: (P_{j,k} + I^{sq}_{j,k}r_j)-P^{inj}_{i,k}, \: \: \: i \in \mathcal{T}, \label{distflow_1} \\
			& Q_{i,k} = \: \sum_{j\in \mathcal{C}_i} \: \: (Q_{j,k} + I^{sq}_{j,k}x_j)-Q^{inj}_{i,k}, \: \: \: i \in \mathcal{T},   \\
			& V^{sq}_{A_i,k} = \: V^{sq}_{i,k} + 2(r_iP_{i,k} + x_{i}Q_{i,k})+  I^{sq}_{i,k}(r^{2}_{i} + x^2_i), \: \: \: i \in \mathcal{U}, \label{distflow_3}\\
			& I^{sq}_{i,k} =\: \frac{P^{2}_{i,k} + Q^2_{i,k}}{V^{sq}_{i,k}} , \:  \: \: \: i \in \mathcal{U} \label{distflow_4} ,
		\end{align}
\end{subequations}
\sloppy where $r_j$ and $x_j$ are the resistance and inductive reactance of the line $j$ and $P_{i,k},Q_{i,k},P^{inj}_{i,k}, Q^{inj}_{i,k},V^{sq}_{i,k}$ and $I^{sq}_{i,k}$ are the $i$-th elements of vectors \(P_k, Q_k, P^{inj}_k\), \(Q^{inj}_k, V^{sq}_k, I^{sq}_k\) respectively. In particular, $P_{i,k}$ and $Q_{i,k}$ are the active and reactive power flow from bus $A_i$ to bus $i$ respectively, $P^{inj}_{i,k}$ and $Q^{inj}_{i,k}$ are active and reactive power injections at bus $i$ respectively.  

In the above branch flow model, equations \eqref{distflow_1}-\eqref{distflow_3} are linear constraints while \eqref{distflow_4} is nonlinear and non-convex in the decision variables. Past work has explored several approaches for obtaining tractable approximations to the non-convex power flow equations.\footnote{ Since the ancillary service requirements imposed by Belgian TSO necessitates introduction of integer variables which considerably increases the computational burden, we resorted to using linearized power flow equations so that the overall problem remains MILP and the equality constraints remain linear. Algorithms for MILP problems are quite mature with solvers such as MOSEK performing extremely well in practice, while MINLP problems are intractable for the most part. In addition, non-convex equality constraints induced by nonlinear power flow equations scale with the size of the network, and will lead to exponential increase in computation time for larger networks. Therefore, we believe that linearization of power flow equations is essential for the problem studied in this work.} Inspired by past works such as \cite{florian_as,yang_estimate}, we have considered a linearization of \eqref{distflow_4} using first order Taylor's series approximation around an initial estimate of $I^{sq}_{i*}(P_{i*},Q_{i*},V^{sq}_{i*})$ given by
\begin{align}
I^{sq}_{i,k} = & I^{sq}_{i*} + (P_{i,k}-P_{i*})\Big[\frac{\partial I^{sq}_{i,k}}{\partial P_{i,k}}\Big]_{I^{sq}_{i*}} \nonumber 
\\ & + (Q_{i,k}-Q_{i*})\Big[\frac{\partial I^{sq}_{i,k}}{\partial Q_{i,k}}\Big]_{I^{sq}_{i*}} \!\! \! \! + (V^{sq}_{i,k}-\!V^{sq}_{i*})\Big[\frac{\partial I^{sq}_{i,k}}{\partial V^{sq}_{i,k}}\Big]_{I^{sq}_{i*}}. \label{taylor}
\end{align}
The solution of the multi-stage optimization problem obtained at the previous iteration is used as the point of linearization for better accuracy.

The branch power flow limits (thermal limits) for each line in the ADN are given in terms of the maximum apparent power handling capability ($S^{max}_i$) of the lines. Assuming a relatively high value of branch power flow limit, the limits on active and reactive power flows can be set as
\begin{subequations}
		\label{branch_max}
		\begin{align}
			& -P^{max}_i \leq P_{i,k} \leq P^{max}_i, \\
			& -Q^{max}_i \leq Q_{i,k} \leq Q^{max}_i,
		\end{align}
\end{subequations}
where $P^{max}_i$  and $Q^{max}_i$ are both set equal to $(S^{max}_i/\sqrt{2})$. This corresponds to an inner approximation of the circular limit by a square contained entirely within the circle. The voltage limit on buses of the ADN are given by
\begin{equation}
		\label{v_max}
		(V^{min})^{2} \leq V^{sq}_{i,k} \leq (V^{max})^2.
\end{equation}


\subsection{Passive Voltage Support Constraints}
	
	\begin{figure}
		\centering
		\includegraphics[width=0.3\textwidth]{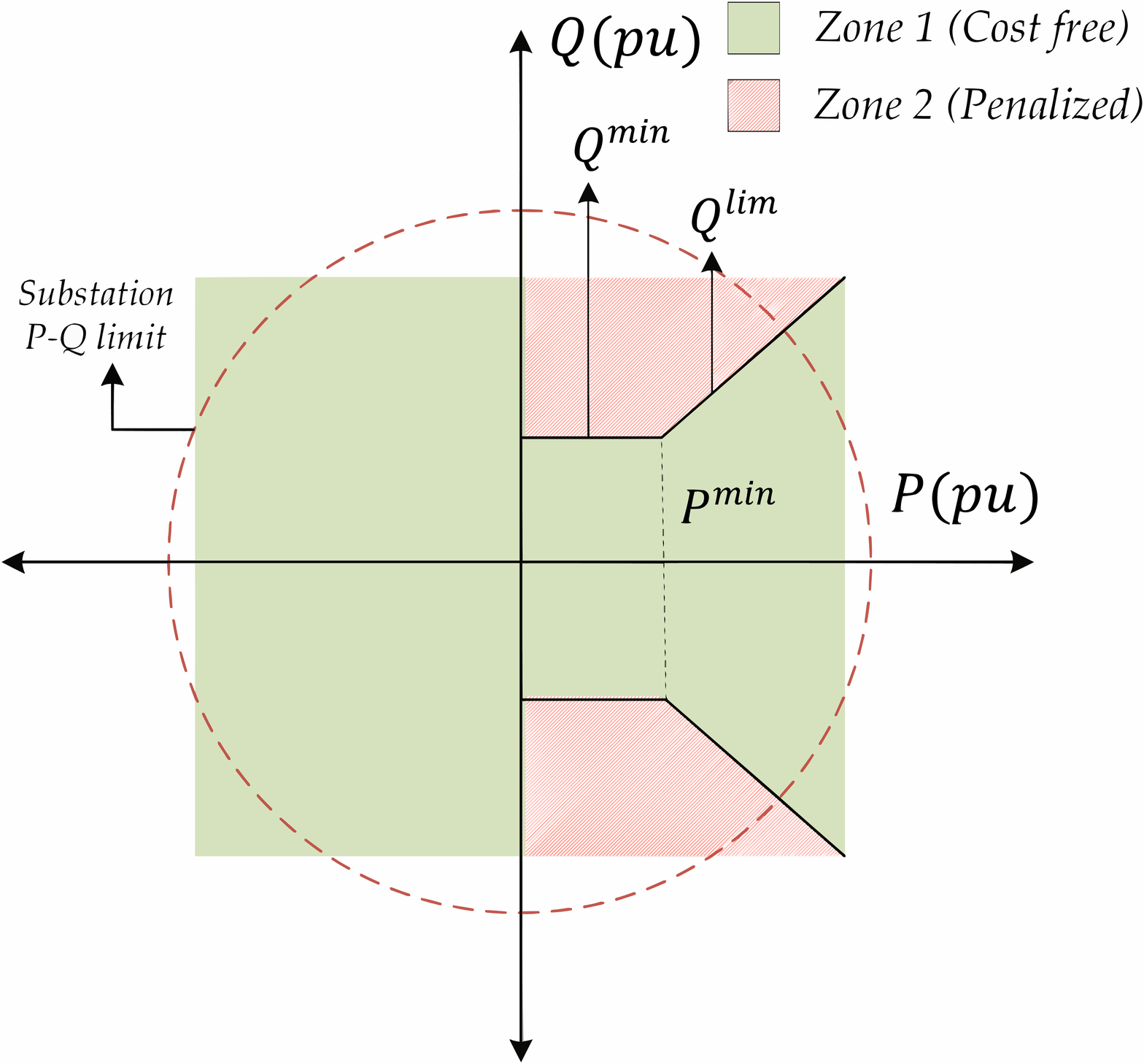}
		\caption{Passive voltage support curve by Belgian's TSO proposal (\textit{Figure is not drawn to scale}).}
		\label{voltage_support}
	\end{figure}
	
	In our study, we have considered the Belgian TSO's proposal \cite{as_belgian} to penalize DSO if the DN is operated beyond the recommended zone, as shown in Fig. \ref{voltage_support}. According to the proposal in \cite{as_belgian}, there are two zones of operation.
	\begin{itemize}
		\item Zone 1 refers to the region with $\cos(\phi) \geq 0.95$ ($\phi$ being the power factor) for $P^{ex}_k \geq P^{min}$ and the rectangular region formed by $P^{min}$ and $Q^{min}$ for $P^{ex}_k < P^{min}$ or when ADN is supplying power to the TN. This zone is the cost free zone and it is desired that the active and reactive power exchange belong to this zone. 
		\item Zone 2 refers to the region with  $\cos(\phi) < 0.95 $ for $P^{ex}_k \geq P^{min}$ and the zone excluding the rectangular zone formed by $P^{min}$ and $Q^{min}$ for $0 \leq P^{ex}_k < P^{min}$. When the power exchange lies in this zone, the ADN is penalized.
	\end{itemize}
	For active power below $P^{min}$, the reactive power is constrained to be below a certain constant $Q^{min}$. The cost to the ADN at time $k$ under the passive voltage support scheme is given by
	\begin{equation}
		\label{as_obj}
		C^{tn}_k := \begin{cases}
			c^p (|Q^{ex}_k| -Q^{lim}), \: & \text{if} \: \: Q^{ex}_k \: \text{belongs to} \: \textit{Zone 2}, \\
			0, \: & \text{if} \: \: Q^{ex}_k \: \text{belongs to} \: \textit{Zone 1},
		\end{cases}
	\end{equation}
	where $c^p$ is the penalty factor, and $Q^{ex}_k$ is the reactive power exchange with the TN. Our aim is to represent the above discontinuous cost function in terms of suitable mixed integer linear constraints that will ensure that the power exchanges lies in Zone 1 (cost free zone) of Fig. \ref{voltage_support}. 
	\begin{remark}
		While the past work \cite{florian_as} also considered the above passive voltage support scheme, their reformulation of the above cost function into mixed integer linear inequality constraints (equation (10) of \cite{florian_as}) is incorrect to the best of our understanding. In particular, considering the value of $Q^{min}$ as given in Table II of \cite{florian_as}, their formulation fails to capture the separating boundary between Zone 1 and Zone 2 for $P^{ex}$ greater than $P^{min}$ in Fig. \ref{voltage_support}. Even if we consider any other value of $Q^{min}$, their formulation fails to capture the rectangular region formed by $P^{min}$ and $Q^{min}$ in Fig. \ref{voltage_support}. 
	\end{remark}
	We now propose an alternative mixed integer reformulation that exactly captures all the intricacies of \eqref{as_obj} shown in Fig. \ref{voltage_support}. We introduce three binary variables:
	\begin{itemize}
		\item $\delta^p_k$ which takes value $1$ when $P^{ex}_k$ is negative and vice versa,
		\item $\delta^\phi_k$ which takes value $1$ when $Q^{ex}_k$ is within Zone 1 and value $0$ when $Q^{ex}_k$ is within Zone 2, and
		\item $\delta_k$ which takes value $0$ when $P^{ex}_k \geq P^{min}$ and vice versa.
	\end{itemize}


Consider the following collection of inequalities where
    \begin{itemize}
        \item the following inequalities
        \begin{subequations}
            \begin{align}\label{as_first}
			& C^{tn}_k \geq 0, \\
			& C^{tn}_k \leq M^p(1-\delta^p_k), \\
			& C^{tn}_k \leq M^p(1-\delta^{\phi}_k); 
            \end{align}
        \end{subequations}
        where $M^p$ is the `Big-M' constant, ensure that the penalty cost for ADN (i.e. $C^{tn}_k$) is bounded by $[0,M^p]$ if operating point $(P^{ex}_k,Q^{ex}_k)$ is in Zone $2$, otherwise it is zero;
        
        \item in the following inequalities
        \begin{subequations}
            \begin{align}
                \label{q_ex_bound}
			& -\big((C^{tn}_k/c^p) + Q^{lim} + (M^p/c^p)(\delta^p_k + \delta^{\phi}_k)\big)\leq Q^{ex}_k \nonumber \\
                & \qquad \leq (C^{tn}_k/c^p) + Q^{lim} + (M^p/c^p)(\delta^p_k + \delta^{\phi}_k), \\
                \label{q_ex_lim1}
			& -(M^p \delta^{\phi}_k -Q^{lim})\leq Q^{ex}_k \leq (M^p \delta^{\phi}_k -Q^{lim}), \\
			\label{q_ex_lim2}
			& -\big(Q^{lim} +  M^p (1-\delta^{\phi}_k) + \delta^p_k M^p\big)\leq Q^{ex}_k \nonumber \\
                & \qquad \leq Q^{lim} +  M^p (1-\delta^{\phi}_k) + \delta^p_k M^p;
            \end{align}
        \end{subequations}
        equation \eqref{q_ex_bound} bounds $C^{tn}_k$ in $[c^p (|Q^{ex}_k| -Q^{lim}), M^p]$ if operating point is in Zone 2, and equations \eqref{q_ex_lim1}- \eqref{q_ex_lim2} ensure that $Q^{ex}_k$ is bounded by $[Q^{lim}, Q^{lim}+M^p]$ if operating point is in Zone 2; 
        \item the following inequalities
        \begin{subequations}\label{p_min}
            \begin{align}
                & -\big(M^p(1-\delta_k)+\zeta\big)\leq P^\mu_k \tan\phi \leq M^p(1-\delta_k)+\zeta, \\
			& P^{ex}_k-P^\mu_k \leq \zeta (1-\delta_k) +\delta_k P^{min}, \\
			& P^{ex}_k-P^\mu_k\geq - \delta_k M^p, \\			
			& P^{\mu}_k \geq (1-\delta_k)P^{min},
            \end{align}
        \end{subequations}
        where $\zeta$ is a very small positive constant, define an auxiliary decision variable $P^{\mu}_k$ which takes value $P^{ex}_k$ if $P^{ex}_k \geq P^{min}$, and $\zeta$, otherwise. The $P^{\mu}_k$ is used to capture the linearly increasing Zone 1 of Fig. \ref{voltage_support};       
        \item the separating boundary between Zone 1 and Zone 2 is formed by the following equality constraint,
        \begin{align}
            & Q^{lim} = \delta_k Q^{min} + P^\mu_k \tan\phi,
        \end{align}
        where $Q^{lim}$ takes value $Q^{min}$ if $P^{ex}_k$ is below the limit of $P^{min}$ otherwise, it increases linearly with $P^{ex}_k$;
        \item and finally the constraints
        \begin{subequations}
            \begin{align}
                & P^{ex}_k \geq -\delta^p_k M^p, \\
			& P^{ex}_k \leq M^p(1-\delta^p_k),
            \end{align}
        \end{subequations}
        ensure that $P^{ex}_k$ is bounded by $[0,M^p]$ or $[-M^p,0]$ depending on its sign.
    \end{itemize}
	
	The reactive and active power limits $Q^{min}$  and $P^{min}$ are defined as
	\begin{subequations}
		\label{tr_bound}
		\begin{align}
			& -\big((V_{tr}/100) S_{tr}\big)\leq Q^{min} \leq (V_{tr}/100) S_{tr}, \\
                \label{as_last}
			& P^{min} = Q^{min} \cot \phi, 
		\end{align}
	\end{subequations}
	where $V_{tr}$ is the substation transformer's short circuit voltage (in \%) and $S_{tr}$ is the nominal apparent power of the transformer. The constraint \eqref{tr_bound} is used to prevent the transformer from becoming disconnected during very lightly loaded condition of DN \cite{stavros_as}.
	 
 	\begin{figure}
		\centering
		\includegraphics[width=0.6\linewidth]{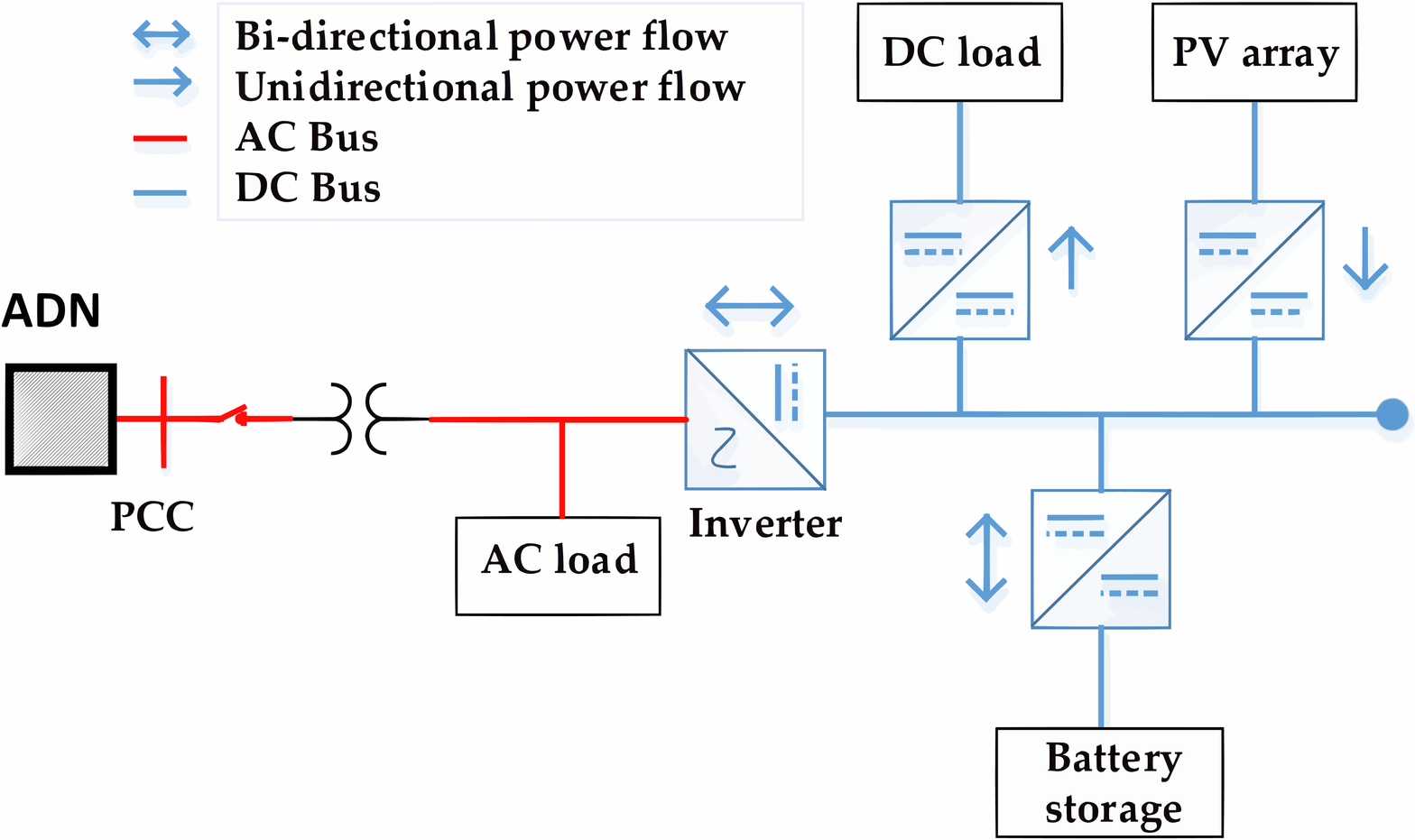}
		\caption{Schematic of microgrid.}
		\label{pic_microgrid}
	\end{figure}

	\subsection{Operational Constraints of Microgrids}
 
	Each microgrid is equipped with an inverter at the point of interface between AC (load, connection with ADN) and DC (storage, load, and PV units) components (see Fig. \ref{pic_microgrid}). Smart inverters now have the capability to provide reactive power support under different modes of operations \cite{ieee_standerd,ieee_report} such as, constant power factor mode, voltage reactive power mode, active power-reactive power mode and constant reactive power mode. The recent development reported in \cite{Q_at_night} makes it possible to use full capability of inverter without any constraint on operating power factors. Such developments enable inverters to provide purely reactive power that can replace the additional installations of reactive power compensators. In a recent work \cite{stavros_as}, the authors suggest to use this mode for providing voltage support based ancillary services. We follow this approach and impose the following constraints on active and reactive power exchange through inverters that interface microgrids to DN: 
	\begin{equation}
		\label{inverter}
		(P^{inv}_{i,k})^2 + (Q^{inv}_{i,k})^2 \leq (S^{inv}_{i})^2, \: \: \: \forall \: i \in \mathcal{M}, 
	\end{equation}
	where the set $\mathcal{M}$ holds the bus indices at which microgrids are located, $P^{inv}_{i,k} := P^{inj}_{i,k} + P^{load,ac}_{i,k}- P^{curt}_{i,k}$ with $P^{load,ac}_{i,k}$ being the local ac load (refer to Fig. \ref{adn_bus}) and $S^{inv}_i$ is the maximum allowable apparent power through the inverter at $i$-th bus. 
	Following \cite{florian_as,yang_estimate}, we consider the following piece-wise linearization of equation \eqref{inverter} and obtain
    \begin{equation}\label{mg_inv}
        a^p_j(P^{inv}_{i,k}) + b^q_j(Q^{inv}_{i,k}) + c_j \leq 0,  \: \: \: \forall \: i \in \mathcal{M},  j\in [l],
    \end{equation}
    where $l$ being the total number of piece-wise segments, and
    \begin{subequations}
        \begin{align}
            & a^p_j= 2\sin(\pi/l) \sin(\frac{\pi}{l}(2j-1)), \\
            & b^q_j= 2\sin(\pi/l) \cos(\frac{\pi}{l}(2j-1)), \\
            & c_j= -(S^{inv}_i \sin(2\pi/l)). 
        \end{align}
    \end{subequations}

Each microgrid considered in our study consists of a Battery Energy Storage System (BESS), RES (PV array generation) and local load as shown in Fig. \ref{pic_microgrid}. The dynamic operation of energy storage systems are represented as
	\begin{subequations}
		\begin{align}
			\label{bat_eq}
			& E_{i,k+1} = E_{i,k} - \eta P^{bat}_{i,k}, \\
			\label{e_inq_1}
			& E_{min} \leq E_{i,k+1} \leq E_{max}, \\
			\label{e_inq_2}
			& P^{bat}_{min} \leq P^{bat}_{i,k} \leq P^{bat}_{max},
		\end{align}
	\end{subequations}
where $k \in [N_p]$, $i \in \mathcal{M}$, $\eta$ is the product of BESS's efficiency and sampling time of the optimization problem. $E_{max}$ and $E_{min}$ are respectively assumed to be $0.9$ and $0.2$ times the battery capacity. The active and reactive power balance equations for microgrid $i \in \mathcal{M}$ at time $k \in [N_p]$ are given by
	\begin{subequations}\label{p_balance}
		\begin{align}
			& P^{inj}_{i,k}=  P^{bat}_{i,k}+ P^{pv}_{i,k} + P^{curt}_{i,k}- P^{load}_{i,k}, \\
			& Q^{inj}_{i,k} = Q^{inv}_{i,k} + P^{curt}_{i,k}\tan\Omega_i - Q^{load}_{i,k} ,
		\end{align}
	\end{subequations}
where $P^{pv}_{i,k}$ is the PV array generation at the $i^{th}$ microgrid, $P^{load}_{i,k}$ is the local load (includes both ac and dc load) of $i$-th microgrid, $Q^{load}_{i,k}$ is the local reactive load of $i$-th microgrid and $\Omega_i$ is load power factor angle at $i$-th microgrid. We assume that reasonable accurate forecast of PV generation and load over the prediction horizon are available to the decision-maker. Note that, load deferment can also be implemented by adding suitable linear constraints as mentioned in \cite{dsm} with minor changes in the proposed framework.  
	
 \subsection{Multi-stage Optimization Problem}
	The objective of the ADN is to minimize the total cost, that includes, 
    \begin{itemize}
        \item the cost of purchasing active power from the TN which equals to $\beta^p_k P^{ex}_k$ with $\beta^p_k$ being the tariff of energy purchase,
        \item the penalty due to violating voltage support constraints (i.e., $C^{tn}_k$), 
        \item the penalty due to resistive loss in the lines of the ADN which is defined as  $\sum_{l=1}^{N_l}\beta^{loss}_l I^{sq}_{l,k} r_l, $ where $\beta^{loss}_l$ is the penalty for energy loss, 
        \item the operational cost of BESS (i.e., $\sum^{N_m}_{j=1}  \beta^b_j P^{bat}_{j,k}$), 
        \item the cost due to load curtailment at microgrids which is defined as $\sum_{j=1}^{N_b}\beta^c_j P^{curt}_{j,k}$, assuming a reasonable value for penalty factor $\beta^c_k$,
    \end{itemize}
    over a prediction horizon of length $N_p$.
    
    Formally, we define
	\begin{align}\label{eq:total_cost}
		C(\mathbf{x}) := & \sum^{N_p}_{k=1} \bigg[\beta^{p}_k P^{ex}_k + C^{tn}_k + \sum^{N_l}_{l=1} \beta^{loss}_l I^{sq}_{l,k} r_l  \nonumber
		\\ & \quad +  \sum_{j=1}^{N_b}\beta^c_j P^{curt}_{j,k} + \sum^{N_m}_{j=1}  \beta^b_j P^{bat}_{j,k} \bigg],
	\end{align}
	where $\mathbf{x} := \{\mathbf{x}_k\}_{k \in [N_p]}$ denotes the decision variables over the entire horizon. The above cost can be stated compactly as $\sum^{N_p}_{k=1} c^{\top}\mathbf{x}_k$, where $c$ is the vector of penalty factors of appropriate dimension. The complete multi-stage optimization problem can now be stated in a compact form as
	\begin{subequations}
		\label{cent_prob}
		\begin{align}
			\min_{\mathbf{x},v} \qquad & \sum^{N_p}_{k=1}  \left[ c^{\top}\mathbf{x}_k \right],\\  
			\label{constraints}
			\text{s.t.} \qquad  & H_1\mathbf{x}_k + V v_k+  h_1 \leq 0, \\
			\label{eq_cons}
			& H_2 \mathbf{x}_k + h_2 =0, \\
			\label{mg_inq}
			& G_1 \mathbf{x}_k + g_1 \leq 0, \\
			\label{mg_x}
			& G_2 \mathbf{x}_k + g_2 = 0,
		\end{align}
	\end{subequations}
	where the constraints hold for all $k \in [N_p]$, $v := \{v_k\}_{k \in [N_p]}$ is the vector of binary variables over the entire horizon, $H_1, H_2, G_1, G_2$ and $V$ are matrices and $h_1, h_2, g_1$ and $g_2$ are vectors of suitable dimensions. Specifically, we encode 
	\begin{itemize}[leftmargin=*]
		\item limits on line flow \eqref{branch_max}, voltage magnitude \eqref{v_max} and AS support constraints \eqref{as_first}-\eqref{as_last} in \eqref{constraints}, 
		\item linearized power flow equations \eqref{distflow_1}-\eqref{distflow_3} and \eqref{taylor} in \eqref{eq_cons},
		\item inequality constraints pertaining to microgrid inverter and BESSs stated in \eqref{mg_inv}, \eqref{e_inq_1} and \eqref{e_inq_2} in \eqref{mg_inq}, and
		\item energy and power balance equations from \eqref{bat_eq} and \eqref{p_balance} in equation \eqref{mg_x}.
	\end{itemize}
	Note that, equations \eqref{constraints} and \eqref{eq_cons} encode constraints pertaining to the entire ADN while equations \eqref{mg_inq} and \eqref{mg_x} represent the constraints that are local to MGs. The above problem is an instance of a mixed-integer linear program (MILP). While MILP problems are inherently NP-hard, the number of integer decision variables in \eqref{cent_prob} is $3N_p$, and does not scale with the size of the network. As a result, it is possible to obtain a globally optimal solution using suitable solvers if $N_p$ is chosen carefully. Further discussions on this issue are presented in Section IV.
	
	\section{The Proposed Distributed Formulation}	
	We now present a distributed formulation of the above multi-stage optimization problem \eqref{cent_prob}. We consider a set of $N_m + 1$ agents; one agent responsible for each microgrid and one agent that corresponds to an ADN operator. We define a connected communication graph $\mathcal{G}:= (\mathcal{V,E})$, where $\mathcal{V} := \{0,1,\dots N_m\}$ is the set of nodes that corresponds to the agents, and $\mathcal{E} \subset \mathcal{V} \times \mathcal{V}$ is the set of communication links $(m,n) \in \mathcal{E}$ implying the agents $m$ and $n$  are connected. While the communication graph can be defined depending on the communication capabilities of the MG nodes, at the very least it is assumed that each MG can communicate with the ADN agent and vice versa. 

 \sloppy
    We now split the decision vector $\mathbf{x}_k$ for the centralized optimization problem \eqref{cent_prob} as follows. We define $\mathbf{z}_{0,k}:= \big[ P_k, Q_k, V^{sq}_k, I^{sq}_k, \!P^{ex}_k, \!Q^{ex}_k, \{\!P^{inj}_{j,k}, \!Q^{inj}_{j,k}, \!P^{curt}_{j,k}\}_{j \notin \mathcal{M}},  C^{tn}_k, v_k \big]$ to be the decision vector for ADN (agent $0$). Similarly, the decision vector for $i$-th MG (agent $i$) is defined as $\mathbf{z}_{i,k}:= \big[P^{bat}_{i,k}, P^{curt}_{i,k}, E_{i,k+1}, Q^{inv}_{i,k} \big]$, where $i \in [N_{m}]$. Note further that it is natural to treat constraints \eqref{mg_inq} and \eqref{mg_x} as internal constraints of MGs and constraints \eqref{constraints} and \eqref{eq_cons} as internal constraints of ADN. However, the variables $\{P^{inj}_{j,k}, Q^{inj}_{j,k}\}_{j \in \mathcal{M}}$ that correspond to active and reactive power exchange between $j$-th MG and ADN are important for both MGs as well as the ADN to satisfy their internal constraints. Therefore, we define $\mathbf{y}_{k}:= \{P^{inj}_{j,k}, Q^{inj}_{j,k} \}_{j \in \mathcal{V}}$ to be shared variables such that each agent maintains a copy of these variables, and the distributed problem is formulated such that agents strive to achieve consensus on these variables in addition to minimizing their local cost. 
    
    The cost for the ADN and the $i$-th microgrid at time $k$ are defined as
        \begin{subequations}
    	\begin{align}
    		c^{\top}_{0,k} \mathbf{z}_{0,k} + \bar{c}^{\top}_{0,k} \mathbf{y}_{0,k} & := C^{tn}_k + \sum_{j \in (\mathcal{T}\setminus \mathcal{M})} \beta^c_jP^{curt}_{j,k} \nonumber
    		\\ & \sum^{N_l}_{l=1} \beta^{loss}_l I^{sq}_{l,k} r_l  +\beta^{p}_k (P^{ex}_k - \sum_{j\in \mathcal{M}} P^{inj}_{j,k}), 
    		\\ c^{\top}_i \mathbf{z}_{i,k} + \bar{c}^{\top}_{i,k} \mathbf{y}_{i,k} & := \beta^c_i P^{curt}_{i,k} + \beta^b_i P^{bat}_{i,k} + \beta^{p}_k P^{inj}_{i,k}.
    	\end{align}
        \end{subequations}

	In other words, the sum of local cost of the ADN and each microgrid over the horizon coincides with the total cost of the centralized problem stated in \eqref{eq:total_cost} when agents reach consensus over the shared variables. The above definition implies the ADN is responsible for minimizing the cost of active power consumption at all nodes that do not have a MG while each MG is responsible for the its own active power consumption cost. Both ADN and the MGs evaluate the active power consumption cost at the price set by the TSO. 
	
	Following similar formulation in \cite{admm_hota}, the problem in equation \eqref{cent_prob} can be equivalently stated as 
	\begin{subequations}
		\label{opt_dist}
		\begin{align}
			\min_{\substack{\mathbf{z}_{0}, \mathbf{y}_{0} \\ \{\mathbf{z}_{i},\mathbf{y}_{i}\}_{i \in [N_m]}}}  \quad & \sum^{N_p}_{k=1} \big[c^{\top}_{0,k} \mathbf{z}_{0,k} + \bar{c}^{\top}_{0,k} \mathbf{y}_{0,k} \\ \nonumber
            & \qquad + \sum^{N_m}_{i=1} ( c^{\top}_i \mathbf{z}_{i,k} + \bar{c}^{\top}_{i,k} \mathbf{y}_{i,k} ) \big],  \\ 
			\label{dn_eq_2}
			\text{s.t.} \qquad \:\: & A_1\mathbf{z}_{0,k} \leq d_1, \\
			\label{dn_eq_1}
			& A_0\mathbf{z}_{0,k} + B_0\mathbf{y}_{0,k} = d_0, 
			\\
			\label{mg_eq}
			& F_i\mathbf{z}_{i,k} + B_i\mathbf{y}_{i,k} \leq f_i, \: \: \: \forall i \in [N_m] \\ 			
			\label{sh_eq1}
			& \mathbf{y}_{m,k} = u_{mn,k}, \: \: \forall m\in \mathcal{V}, \: (m,n) \in \mathcal{E}, \\
			\label{sh_eq2}
			& \mathbf{y}_{n,k} = u_{mn,k}, \: \: \forall n\in \mathcal{V}, \: (m,n) \in \mathcal{E},
		\end{align}
	\end{subequations}
	where the constraints hold for every $k \in [N_p]$. We include the constraints in \eqref{constraints} in terms of equation \eqref{dn_eq_2}, equation \eqref{dn_eq_1} corresponds to equation \eqref{eq_cons}, equations \eqref{mg_inq} and \eqref{mg_x} are merged into equation \eqref{mg_eq}, and auxiliary  variables $u_{mn,k}$ enforce consensus constraints over shared variables between two neighboring agents $m$ and $n$ of the communication graph. 
	
	We leverage the fully distributed ADMM algorithm from \cite{makhdoumi2017convergence,admm_hota} to solve the above problem stated in \eqref{opt_dist}. The complete algorithm is stated in Algorithm \ref{algo}.  We denote by $\lambda_{i,k} \in \mathbb{R}^{p} $ the Lagrange multipliers corresponding to the equality constraints in \eqref{sh_eq1}-\eqref{sh_eq2}. The optimization sub-problem for the ADN using the augmented Lagrangian function is defined as
	\begin{subequations}\label{eq:ADN_ADMM}
		\begin{align}
			\nonumber \min_{\mathbf{z}_{0},\mathbf{y}_{0}} \: & L_{\rho}(\mathbf{z}_{0},\mathbf{y}_{0},\lambda_{0},\{\mathbf{y}_{m}\}_{m \in \mathcal{N}_0})=   \sum^{N_p}_{k=1}   \big[c^{\top}_{0,k} \mathbf{z}_{0,k} + \bar{c}^{\top}_{0,k} \mathbf{y}_{0,k}  \\ \label{sub_1} 
			& +  \mathbf{y}^{\top}_{0,k}\lambda_{0,k} + (\rho/2) \sum_{m \in \mathcal{N}_0} \left\lVert \mathbf{y}_{0,k} - \frac{\widehat{\mathbf{y}}_{0,k}+\mathbf{y}_{m,k}}{2} \right\rVert^2_2 \big],  \\
			\text{s.t.} \quad  & A_0\mathbf{z}_{0,k} + B_0\mathbf{y}_{0,k} = d_0, \\
			& A_1\mathbf{z}_{0,k} \leq d_1, \qquad \forall k \in [N_p],
		\end{align}
	\end{subequations}
	where the ADN agent uses the shared variables $\mathbf{y}_{m,k}$ received from its neighbors $\mathcal{N}_0$ and its local copy of the shared variables obtained in the previous iteration $\widehat{\mathbf{y}}_{0,k}$ in the second term of the augmented Lagrangian to achieve consensus with its neighbors. The hyper-parameter $\rho$ controls the relative weight of the consensus term and the cost terms in the cost function of \eqref{eq:ADN_ADMM}. The Lagrange multipliers are updated locally by the ADN agent as shown in Line 4 of Algorithm \ref{algo}. The above problem is an instance of a mixed-integer quadratic program.  
 
    We now state the subproblem for $i$-th MG below.
	\begin{subequations}
		\label{subproblem_mg}
		\begin{align}
			\nonumber \min_{\mathbf{z}_{i},\mathbf{y}_{i}} \qquad & L_{\rho}(\mathbf{z}_{i},\mathbf{y}_{i},\lambda_{i},\{\mathbf{y}_{m}\}_{m \in \mathcal{N}_i})= \:  \sum^{N_p}_{k=1} \: \:  \big[c^{\top}_i \mathbf{z}_{i,k} + \bar{c}^{\top}_{i,k} \mathbf{y}_{i,k} \nonumber \\  \label{sub_mg_1}
			&  +  \mathbf{y}^{\top}_{i,k}\lambda_{i,k} + (\rho/2)  \sum_{m \in \mathcal{N}_i} \left\lVert \mathbf{y}_{i,k} - \frac{\widehat{\mathbf{y}}_{i,k}+\mathbf{y}_{m,k}}{2} \right\rVert^2_2  \big],  \\
			\text{s.t.} \qquad & F_i\mathbf{z}_{i,k} + B_i\mathbf{y}_{i,k} \leq f_i, \: \: \: \forall k \in [N_p],  
		\end{align}
	\end{subequations}
	where $\mathcal{N}_i$ is the set of neighbors for $i$-th agent, and $\rho$ is the penalty factor for the quadratic regularizer term. The above problem is an instance of a quadratic program.
	
        \begin{figure}
		\centering
		\includegraphics[width=0.8\linewidth]{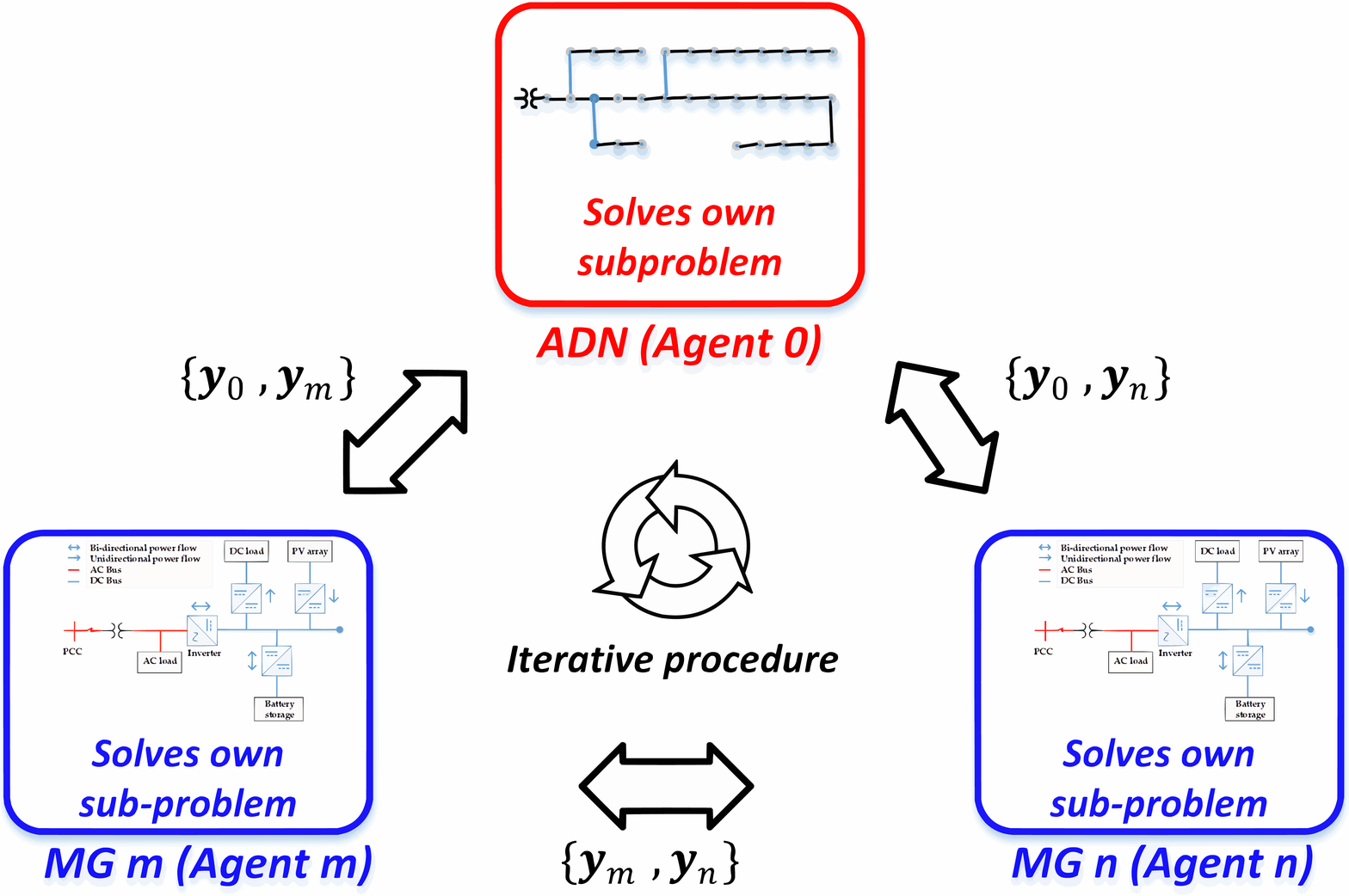}
		\caption{Schematic of distributed operation.}
		\label{pic_distributed}
	\end{figure}
	
	Each agent solves its own sub-problem, and communicates the individual copy of the shared variable vector $\mathbf{y}_k$ with every other agent at each iteration $t$ following the ADMM algorithm. The distributed operation is depicted in Fig. \ref{pic_distributed} where microgrid agents $m\in \mathcal{V}\setminus\{0\}$, and $n\in \mathcal{V}\setminus\{0\}$ with all possible pairs of $(m,n)\in\mathcal{E}$, and $m \neq n$ share their copy of shared variable vectors ($\mathbf{y}_{m}$ and $\mathbf{y}_{n}$ respectively) between each other as well as with the ADN agent. The ADMM generates primal-dual sequences as $\{\mathbf{z}_{j,k},\mathbf{y}_{j,k}\}$, $\{\lambda_{j,k}\}$ that get updated with every iteration $t$ following the Algorithm \ref{algo}. The algorithm terminates when the local copy of the shared variable maintained by the agents becomes approximately equal to the average of the shared variables received from their neighbors up to a tolerance parameter $\epsilon$ (see Line 2 in the Algorithm). 

	\begin{algorithm}\label{algo}
		\SetKwInOut{Input}{Input}
		\SetKwInOut{Output}{Output}
		
		\textbf{Initialization:} At $t=0$, $\mathbf{y}^{t}_{j}=y_0 \in \mathbb{R}^p$ and $\lambda^{t}_{j}=0$ for any $j\in \mathcal{V}$,
		
		\While{$\max_{j \in \mathcal{V}} \left\lVert \mathbf{y}^{t}_{j} - \frac{1}{\lvert\mathcal{N}_j\rvert} \sum_{l \in \mathcal{N}_j} [\mathbf{y}^{t}_{l}] \right\rVert^2_2 \geq \epsilon$}{  
			\For{$j \in|\mathcal{V}|$}{   
				$\lambda^{t+1}_{j} \leftarrow \lambda^t_{j} + \rho \sum_{m\in \mathcal{N}_j} (\mathbf{y}^{t}_{j}-\mathbf{y}^t_{m})$
				
				$(\mathbf{z}^{t+1}_{j},\mathbf{y}^{t+1}_{j}) \leftarrow \arg\min_{(\mathbf{z}^t_{j},\mathbf{y}^t_{j})\in \mathcal{S}}  L_{\rho}(\mathbf{z}^t_{j},\mathbf{y}^t_{j},\lambda^{t+1}_{j},\{\mathbf{y}^t_{l}\}_{l \in \mathcal{N}_j}) $
				
				communicate $\mathbf{y}^{t+1}_{j}$ to neighbors in $\mathcal{N}_j$
				\\ set $\widehat{\mathbf{y}}_{j} \leftarrow \mathbf{y}^{t+1}_{j}$
			}
			$t \leftarrow t+1$
		}
		\textbf{Output:} $\{\mathbf{y}^{t}_{j},\mathbf{z}^t_{j}\}$ for any $j \in \mathcal{V}$.
		\caption{Fully distributed ADMM}
		\label{algo}
	\end{algorithm}

\begin{remark}
The algorithm presented above has been demonstrated to converge to the centralized optimal solution in convex finite-sum settings in \cite{makhdoumi2017convergence}. However, the sub-problem of the ADN agent in \eqref{eq:ADN_ADMM} takes the form of mixed integer quadratic programming, for which theoretical convergence guarantees are not known. Nevertheless, convergence of ADMM-based algorithms, where the centralized problem is a MILP, have been empirically observed in power systems applications \cite{shen2020hierarchical,jian2019hierarchical,dvorkin2018consensus}. In the following section, we present a thorough empirical analysis of the convergence of Algorithm \ref{algo} for different choice of hyperparameters and network sizes.
\end{remark}

\begin{remark}
		In contrast with other ADMM based distributed optimization formulations, such as the ones studied in \cite{peng2016distributed,admm_dual_a, admm_dual_b}, the above approach requires agents to only communicate a subset of the primal decision variables that are relevant for other agents. Since the primal decision variables correspond to active and reactive power signals, a malicious agent can not easily tamper with it and lead the distributed scheme to arrive at a solution that is favorable to itself instead of the social optimal solution. In our scheme, dual variables ($\lambda^t_{j,k}$), i.e., shadow price signals, are locally updated by each agent and not shared among neighbors. 
	\end{remark}

 \section{Numerical Results}
 
	\subsection{Simulation Setup} 
\begin{figure}
		\centering
		\includegraphics[width=0.8\linewidth]{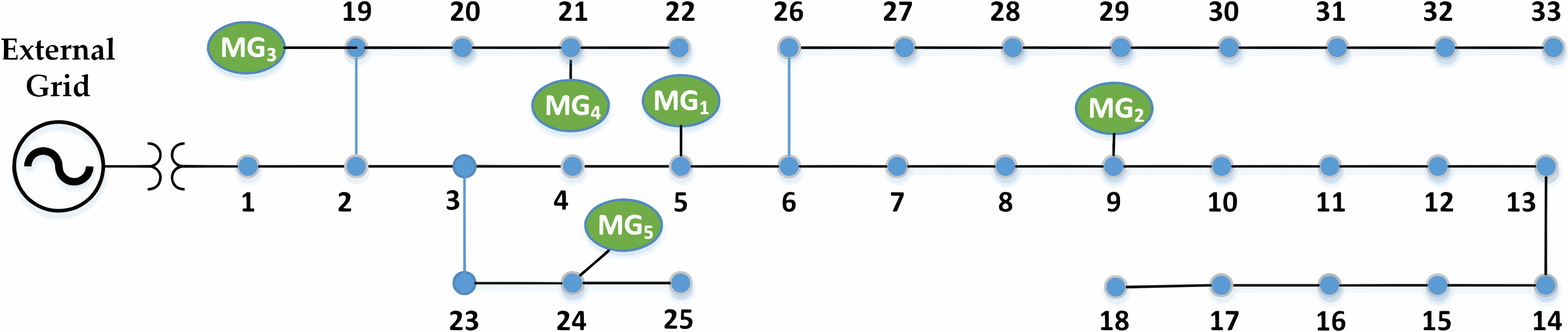}
		\caption{Modified IEEE 33-bus distribution network}
		\label{adn_bus}
	\end{figure}

\begin{figure*}
		\centering
		
		\begin{subfigure}{0.3\linewidth}
			\centering
			\includegraphics[width=\linewidth]{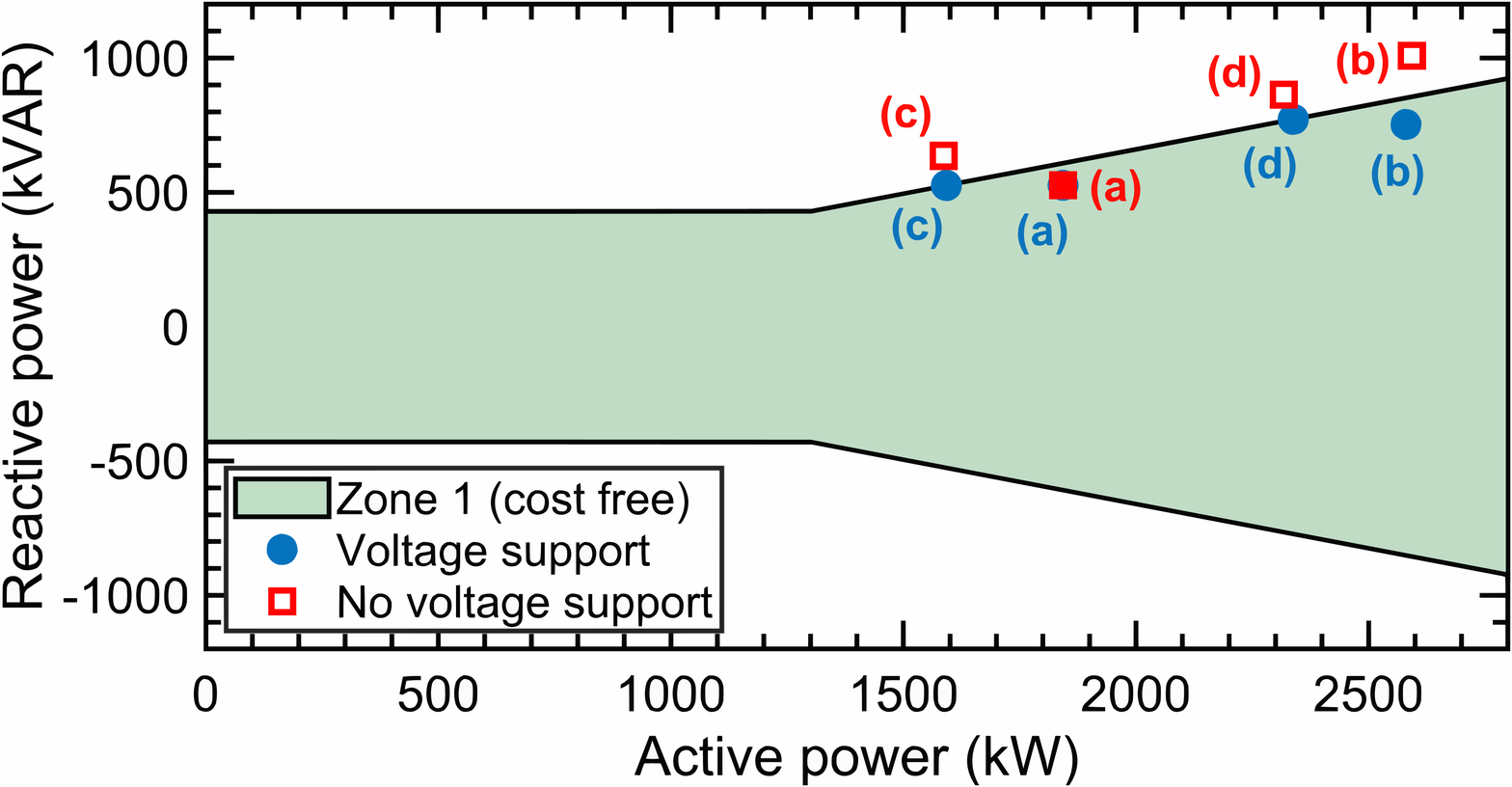}
			\caption{}
			\label{pic_as}
		\end{subfigure}
		\begin{subfigure}{0.3\linewidth}
			\centering
			\includegraphics[width=\linewidth]{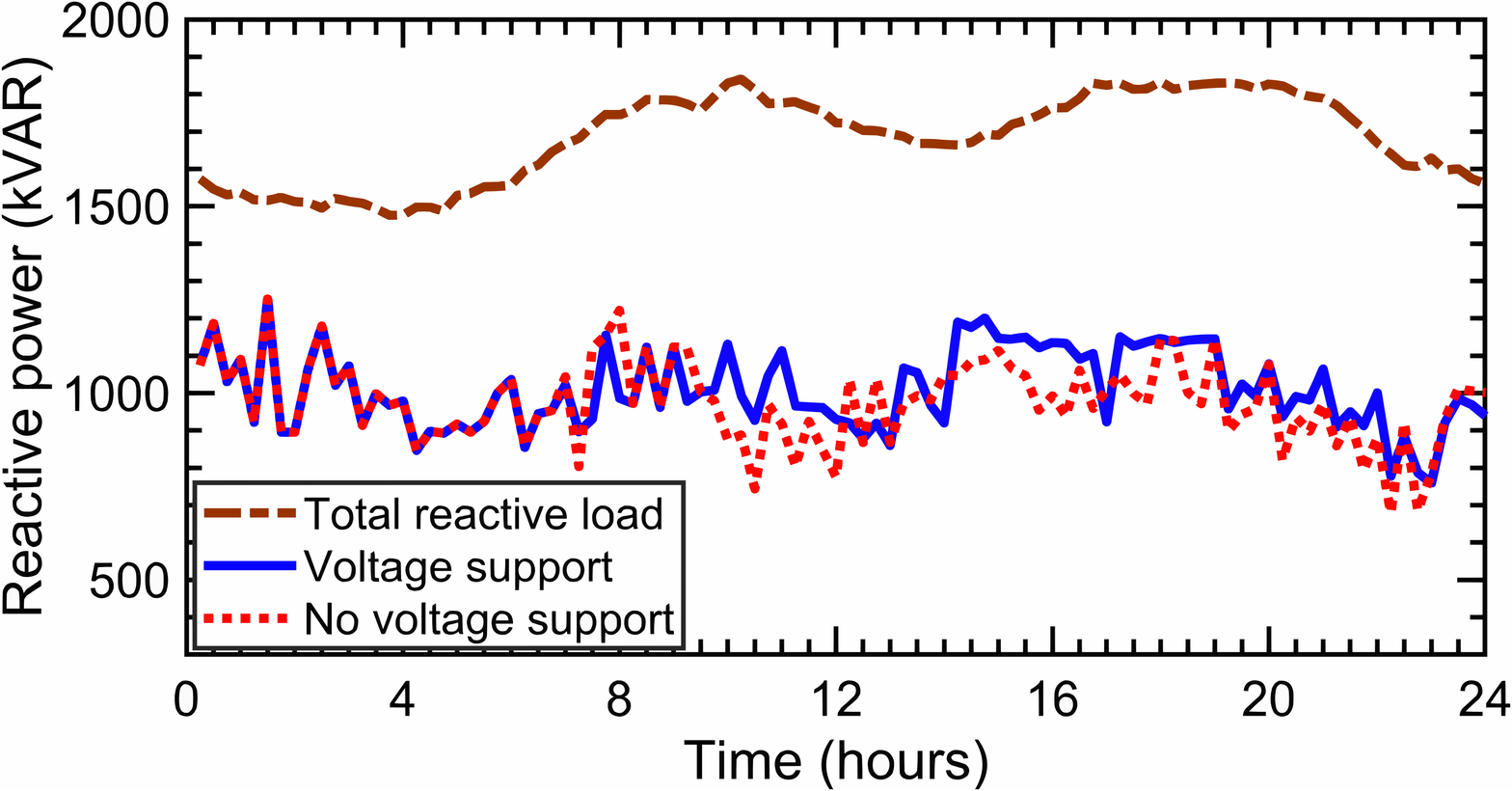 }
			\caption{}
			\label{pic_inv}
		\end{subfigure}
		\begin{subfigure}{0.3\linewidth}
			\centering
			\includegraphics[width=\linewidth]{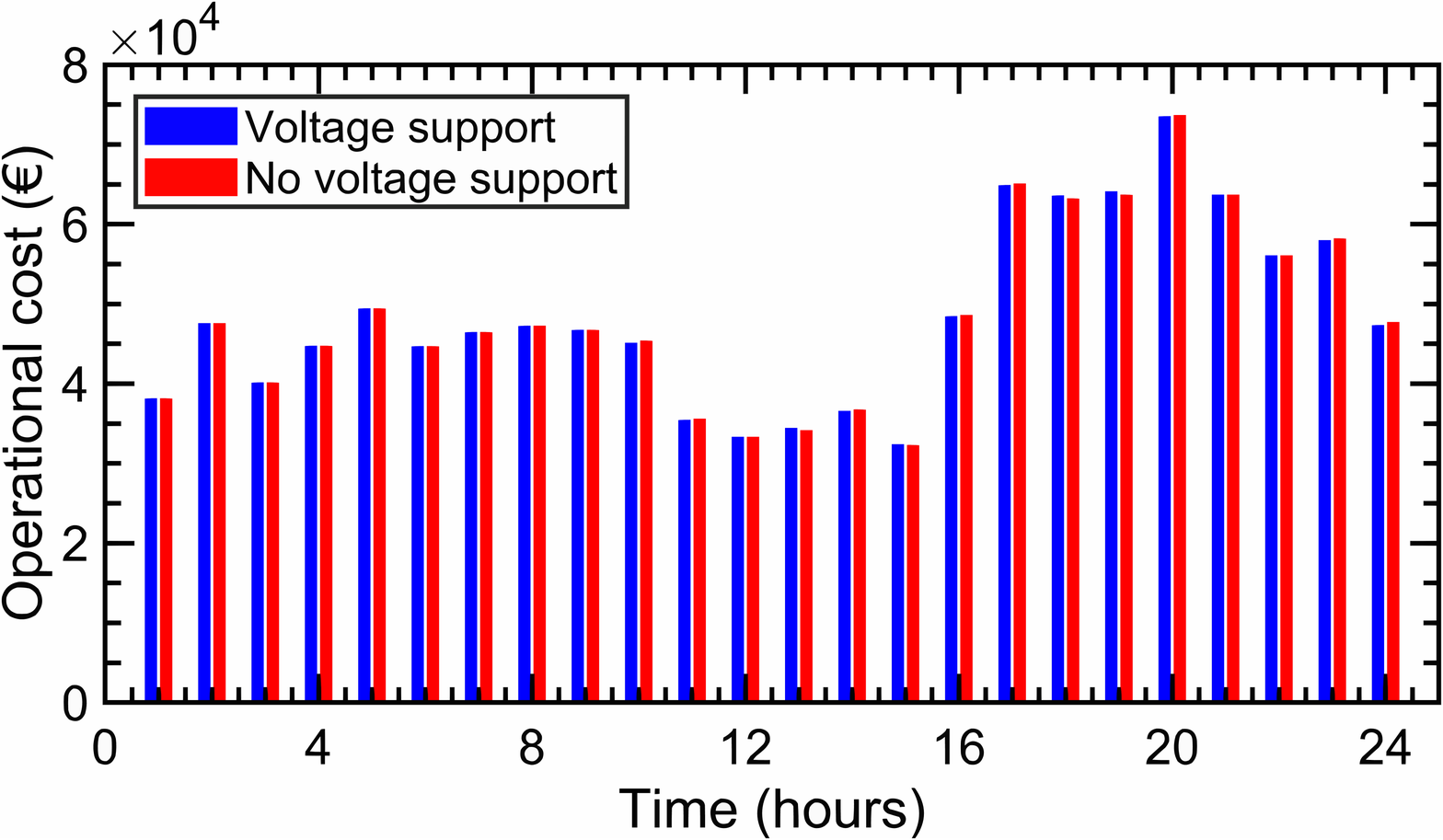}
			\caption{}
			\label{pic_cost}
		\end{subfigure}

		\begin{subfigure}{0.3\linewidth}
			\centering
			\includegraphics[width=\linewidth]{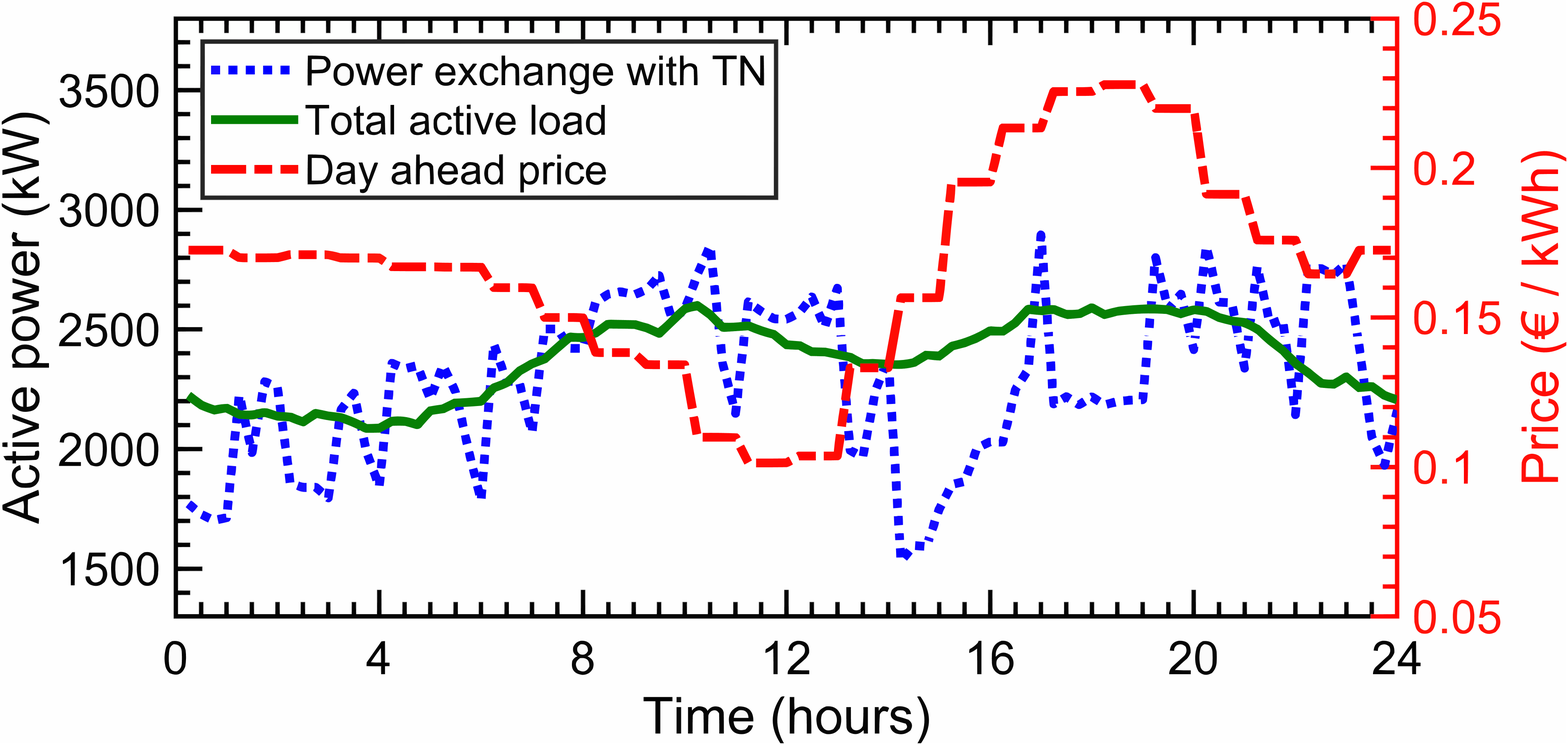}
			\caption{}
			\label{pic_exchange}
		\end{subfigure}
		\begin{subfigure}{0.3\linewidth}
			\centering
			\includegraphics[width=\linewidth]{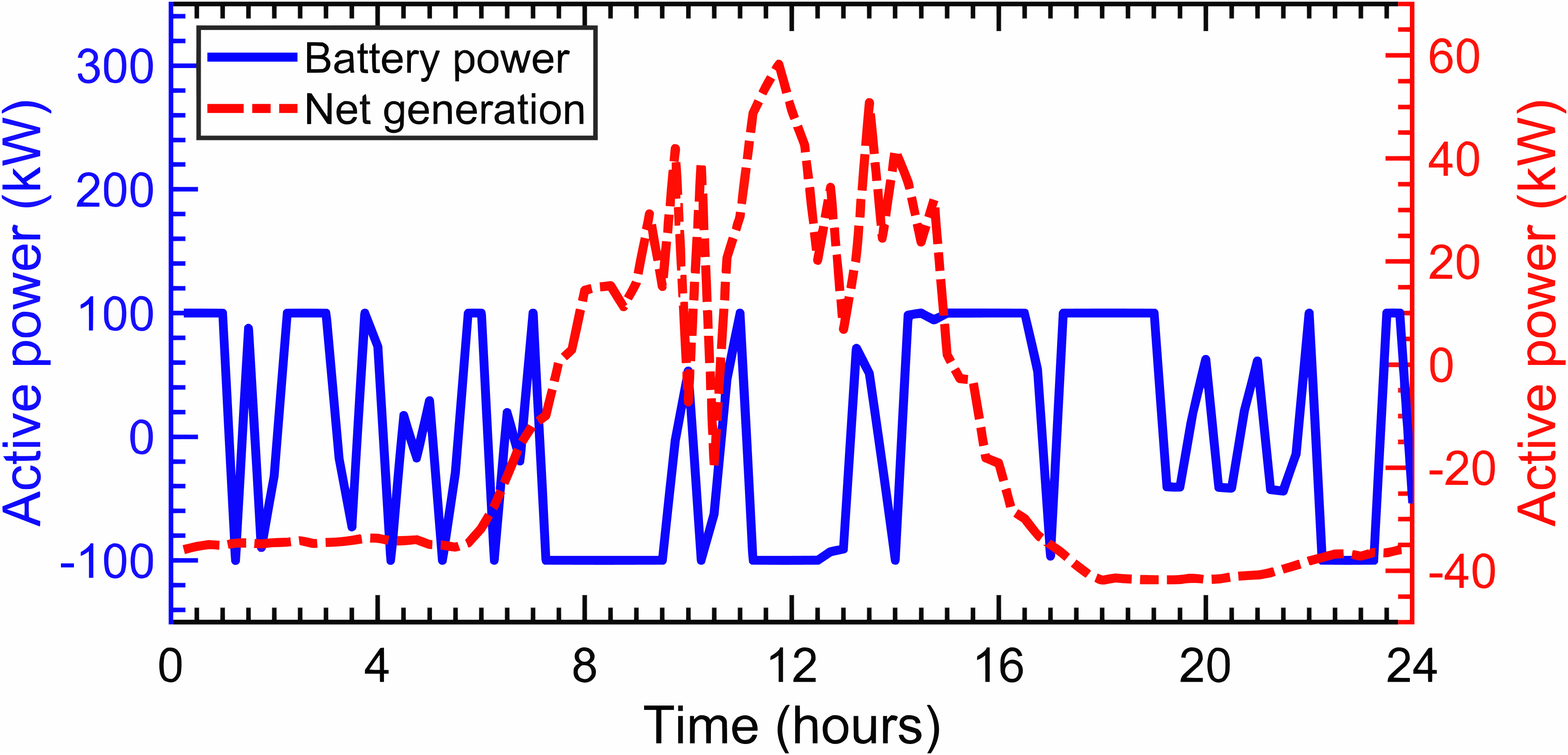 }
			\caption{}
			\label{pic_battery}
		\end{subfigure}   
		\begin{subfigure}{0.3\linewidth}
			\centering
			\includegraphics[width=\linewidth]{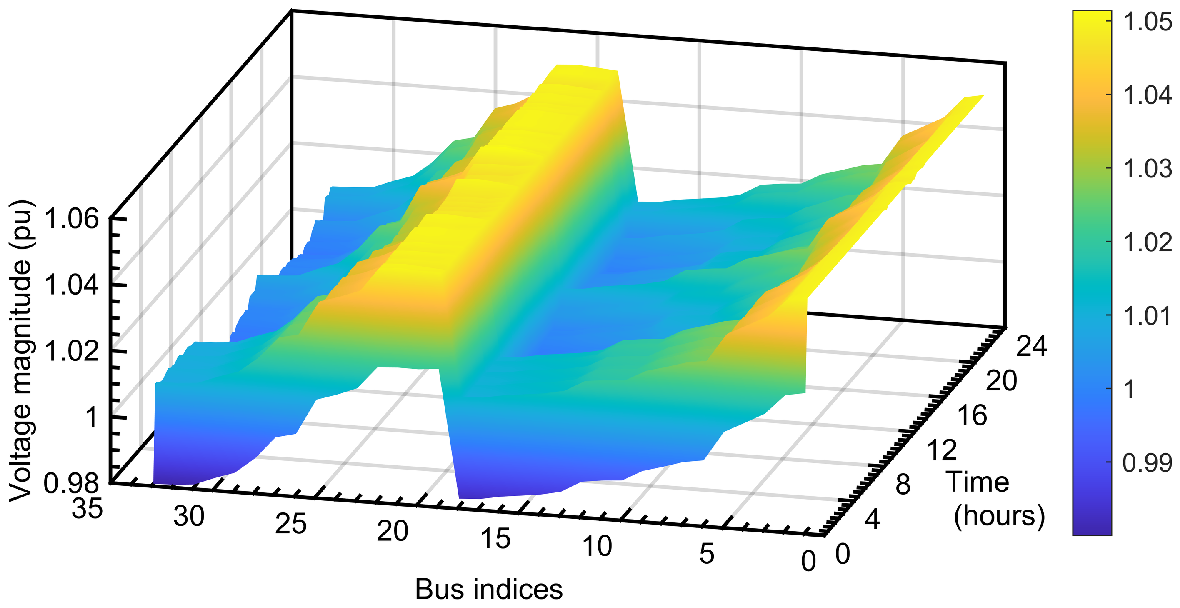}
			\caption{}
			\label{pic_voltages}
		\end{subfigure}  
		\caption{(a)  Passive voltage support curve,  (b)  total reactive load of DN and aggregated reactive power generation by all inverters both with and without voltage support AS constraints, (c) comparison of operational cost (total operational cost under voltage support is 1162460.52 \(\text{\euro}\) and without voltage support is 1162927.08 \(\text{\euro}\) ), (d) active power exchange between TN and DN, total active load of DN and day ahead price by Belgian TSO \cite{entsoe}, (e) net generation (PV generation subtracted by local load) in MG 1 and charging/ discharging power of BESS (positive value implies discharging and vice versa), (f) true bus voltages (solution retrieved from Newton-Raphson load flow) throughout 24-hour operation.}
	\end{figure*}

To validate our proposed formulation, a modified IEEE 33-bus \cite{baran1989network} DN (Fig. \ref{adn_bus}) is considered for numerical simulation. Five microgrids are added to nodes $5,9,19,21$ and $24$ respectively. Each microgrid is equipped with local PV generation, local load, storage and inverter as shown in Fig. \ref{pic_microgrid}. Each agent (the ADN as well as the microgrids) is assumed to be able to communicate with every other agent (i.e., the underlying communication graph is a complete graph). The necessary simulation parameters and system information are provided in Table \ref{table_parameter} and are chosen to simulate the desired behavior of the agents in the scope of the optimization (as in \cite{florian_as}). The cost and penalty parameters are chosen to maintain consistency with the peak tariff rate shown in Fig. \ref{pic_exchange}. 
 
	\begin{table}[t]
		\centering
		\caption{System information and simulation parameters}
		\begin{tabular}{|p{3 cm}|c|c|}
			\hline
			\textbf{Parameter}  & \textbf{Symbol} & \textbf{Value}  \\
			\hline
			Prediction horizon  & $N_p$           & $10$            \\
			\hline
			Rated DN voltage    &     $V_{rated}$    		   & $12.66$ kV		  \\
			\hline
			BESS's capacity       &      -       &    600 kWhr          \\ 
			\hline
			Coefficient of BESS's dynamic model       &      $\eta$        &    0.225 hr          \\ 
			\hline
			PV plant rated power&		$P_{pv,peak}$		   & $400$ kW         \\
			\hline
			Branch flow limit &		$S^{max}$		   & $1200$ kVA         \\
			\hline
			Active power limit in Fig. \ref{voltage_support}					 & $P^{min}$ 	   & $0.5 \times$ (peak exchange) \\
			\hline
			Reactive power limit in Fig. \ref{voltage_support}    				 & $Q^{min}$       & $0.33 P^{min}$       \\ 
			\hline
			Inverters capacity  &		$S^{inv}$		   & $250$ kVA        \\
			\hline
			Maximum charging/ discharging power of BESS                    & $P^{bat}_{max/min}$ & $(+/-)100$ kW  \\
			\hline
			Piece-wise segments in \eqref{mg_inv} & $l$  & $16$              \\
			\hline
			Large constant of Big-M method                   & $M^p$          & $10000$            \\
			\hline 
			Cost of BESS power utilization  & $\beta^{b}$          & $0.1519$ \text{\euro}/kWh \\
			\hline 
			Penalty due to load curtailment  & $\beta^{c}$          & $0.506$           \text{\euro}/kWh  \\
			\hline 
			Penalty due to line loss & $\beta^{loss}$          & $0.075$   \text{\euro}/kWh         \\
			\hline
			Penalty in zone 2 of Fig. \ref{voltage_support} & $c^p$          & $5$   \text{\euro}/kVAR         \\
			\hline 
			Power factor of AC load in $i$-th microgrid & $\cos(\Omega_i)$          & $0.8$   \\
			\hline
		\end{tabular}
		\label{table_parameter}
	\end{table}

For the optimization problem, Python based package Pyomo \cite{pyomo} is used as modeling language and MOSEK \cite{mosek} is used as a solver with default settings. Numerical simulation is performed on a Desktop Computer with 2.90 GHz Intel core-i7 processor, and 64-GB RAM configurations. We choose a sampling time of $15$ minutes and solve each multi-stage distributed optimization problem (\eqref{eq:ADN_ADMM} and \eqref{subproblem_mg}) following Algorithm \ref{algo} over an entire day in a receding horizon manner (popularly known as Model Predictive Control) \cite{borrelli2017predictive} with prediction horizon $N_p = 10$. 

 The primary motivation for considering a receding horizon approach is the presence of storage systems in the ADN. If we only optimize the instantaneous cost, then the optimal solution will prioritize use of stored energy (which is cheap), while disregarding its future availability. Such an outcome is unacceptable since it is pertinent to store energy during intervals of excess PV generation and low electricity price so as to make use of them during intervals of power scarcity or high price. As a result, formulating a multi-stage optimization problem is necessary. Within the class of multi-stage optimization, a receding horizon approach allows us to choose a potentially smaller prediction horizon, and repeatedly refine our solution in an online manner as uncertain parameters such as PV generation and load get realized. A smaller $N_p$ also reduces the computational requirement in a considerable manner (compared to a day ahead formulation) since the number of integer variables scales linearly with $N_p$. 
 
\vspace{-10pt}

	\subsection{Provisioning of Passive Voltage Support AS Scheme}
	As mentioned in Section II, we have considered passive voltage support scheme according to the guidelines given by Belgian TSO. For this simulation, we have considered the load profile data of Belgium area and day-ahead price by Belgian TSO \cite{entsoe} for $22^{nd}$ May, $2022$ as shown in Fig. \ref{pic_exchange}. In Fig. \ref{pic_exchange}, the active power exchange with TN is shown under the implementation of passive voltage support scheme. 
	
	The effectiveness of the proposed distributed scheme in providing passive voltage support is shown in Fig. \ref{pic_as} at four critical time stamps of 24-hour operation period. Time stamp (a) refers to early morning (around 04:00 Hours) when ADN is lightly loaded, time stamp (b) refers to morning (around 10:00 Hours) when system load is ramping up to peak, time stamp (c) refers to afternoon (around 15:00 Hours) when there is relatively less load on the system and lastly, time stamp (d) refers to evening (around 21:00 Hours) when the load is ramping down from peak. In all of the time zones, our formulation for passive voltage support scheme ensures that active and reactive power exchange with TN remains in penalty free Zone 1. In contrast, the points under `No voltage support' tag in Fig. \ref{pic_as} move to Zone 2 for time stamps (b), (c) and (d) when the passive voltage support constraints are absent despite the inverters being operated with full capability.
	
	The amount of reactive power drawn from TN is significantly smaller compared to the reactive load consumption by ADN. This is due to the fact that we have exploited the full capabilities of inverters, as mentioned in Section II, and these inverters aid in provisioning of ancillary services by supplying additional reactive power for those time stamps where system operation might have shifted to Zone 2. For instance, in the absence of passive voltage support constraints, Fig. \ref{pic_inv} shows that inverters provide relatively lower amount of reactive power at time stamps (b), (c) and (d) leading to greater demand from TN and consequent operation in Zone 2. When passive voltage support constraints are introduced in the optimization problem, a smaller amount of reactive power is drawn from the grid leading to system operation in the penalty free Zone 1 (as shown in Fig. \ref{pic_as}). This is corroborated in Fig. \ref{pic_inv} which shows that a larger amount of reactive power was supplied by the inverters at those time periods.

	In terms of operational cost, we observe in Fig. \ref{pic_cost} that, compliance of passive voltage support scheme yields almost same daily operational cost as the `No voltage support' counterpart. In particular, the calculation of operational cost includes a) the cost of purchasing active power from TSO using the time-varying tariff rates (shown in Fig. \ref{pic_exchange}), b) the cost of load curtailment, c) cost of renewable energy from PV plants in MGs. Operational cost of BESSs is not considered as it is negligible for 24-hour operation. In order to obtain a fair comparison, cost due to violation of voltage support constraints is not included since this term was not present in the optimization problem of the `No voltage support' counterpart. However, the net incurred penalty due to operation in Zone 2 under `No voltage support' counterpart comes to $6,89,514.2$ \text{\euro} (considering the value of $c^p$ in Table I borrowed from \cite{florian_as}), while the penalty under voltage support constraints is $0$ under our proposed formulation. These results show the feasibility of implementing such voltage support schemes without incurring much additional operational cost while entirely avoiding penalty due to operation in Zone 2. Fig. \ref{pic_battery} shows that BESSs draw power from ADN while the price is relatively low and discharge as the price increases to compensate internal load consumption of MG and ADN. 
	
	\subsubsection*{\textbf{Bus voltage profiles}}
	Due to the linearization of non-linear constraint \eqref{distflow_4} into \eqref{taylor}, the true voltage magnitudes of nodes may deviate from voltages retrieved as solution of the distributed algorithm. Therefore, at every interval, voltage magnitudes are calculated by conducting Newton-Raphson Load flow which computes voltages using the power injections at the nodes for that particular interval. These true voltages are plotted in Fig. \ref{pic_voltages} for all the buses over the $24$ hour operation, and are found to be within $\pm 5\%$ of the rated value.
	
	\subsection{Comparison of Our Proposed Method with \cite{florian_as}}
	
\begin{figure}
		\centering
		\includegraphics[width=0.8\linewidth]{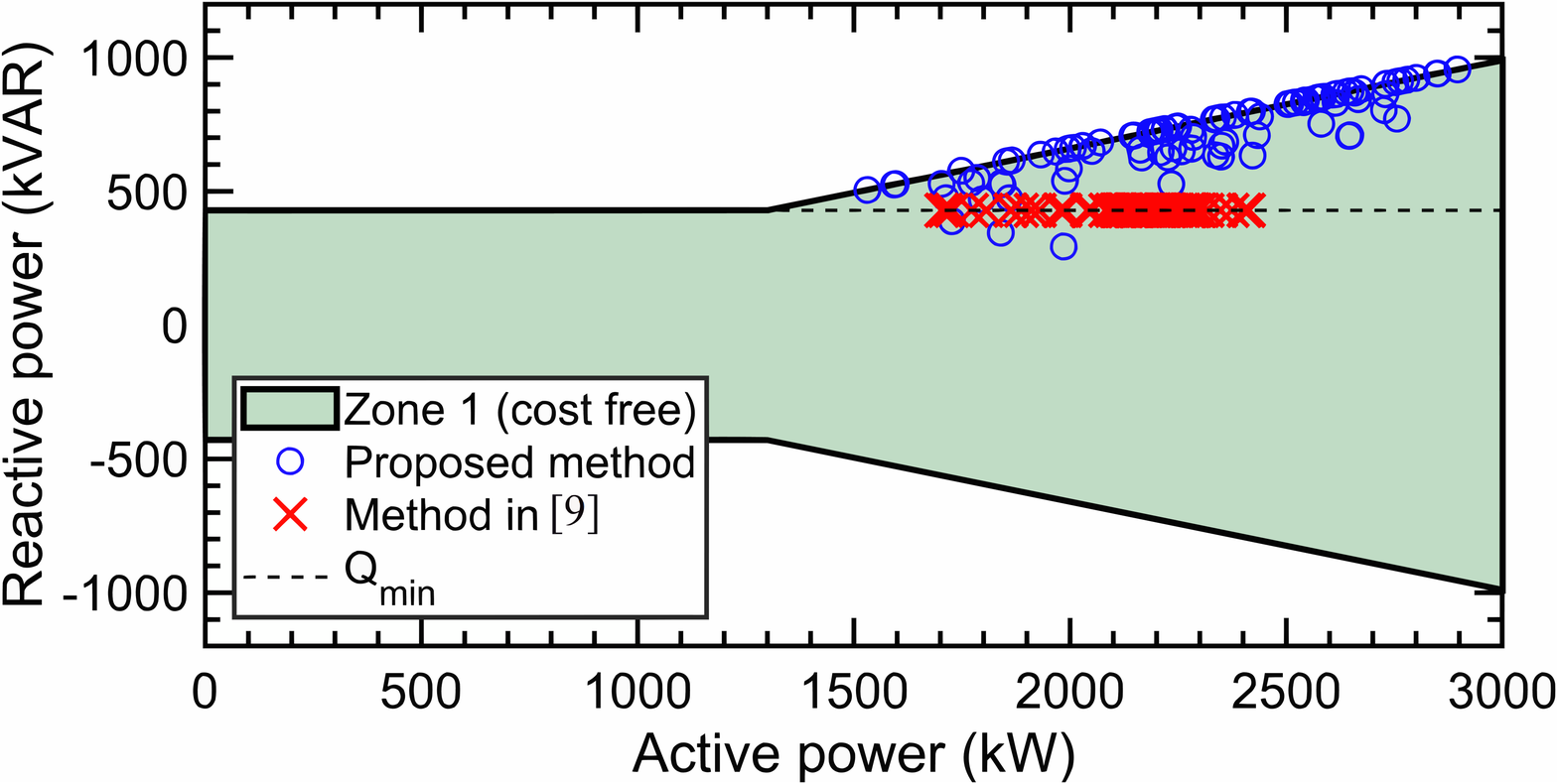}
		\caption{Comparison with \cite{florian_as}.}
		\label{pic_comparison}
	\end{figure}

	In Fig. \ref{pic_comparison}, we compare the solutions obtained under our proposed formulation of the passive voltage support scheme with the formulation presented in \cite{florian_as}. It is evident from the figure that under the constraints presented in \cite{florian_as}, the reactive power exchange at the optimal solutions are restricted to $Q_{min}$ as explained in Remark 1 of Section II. Furthermore, in order to guarantee penalty free operation, the optimal solution obtained under the formulation given in \cite{florian_as} needs to increase load curtailment by $5.31 \%$ of the total load compared to no curtailment of load observed under our proposed scheme. 

	\subsection{Convergence Behavior of the Distributed Algorithm}
	We now examine the convergence behavior of the distributed optimization formulation and provide detailed insights into how to choose different hyper-parameters while deploying such schemes in practice. The earlier test bed, i.e., the modified IEEE 33-bus ADN with five MGs (as shown in Fig. \ref{adn_bus}) is chosen for this study as well.  Prediction horizon for each sub problem is kept at $10$.

\begin{figure}[t]
	\centering
	\begin{subfigure}{0.8\linewidth}
		\centering
		\includegraphics[width=\linewidth]{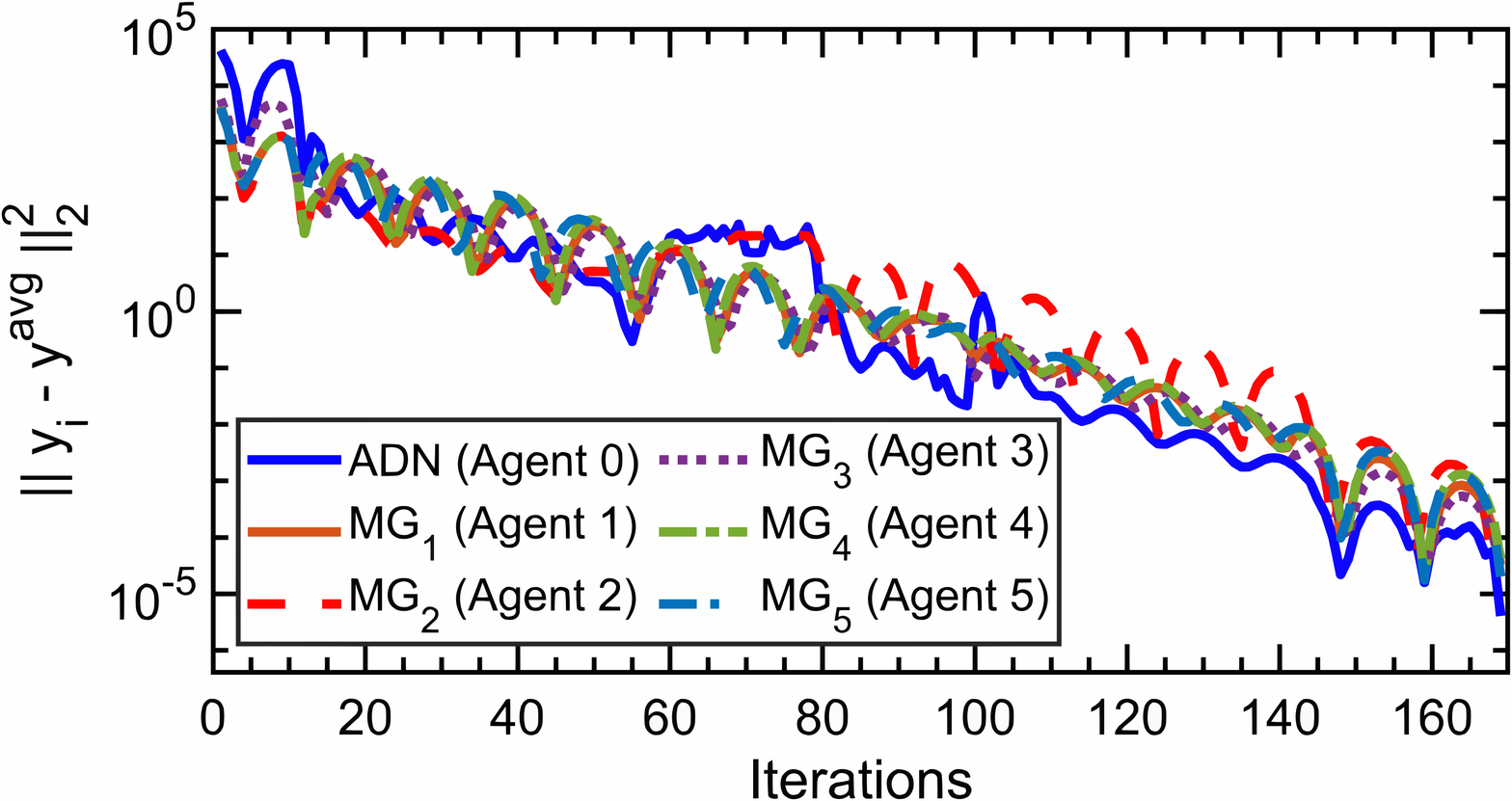}
		\caption{}
		\label{pic_con}
	\end{subfigure}
	
	\begin{subfigure}{0.8\linewidth}
		\centering
		\includegraphics[width=\linewidth]{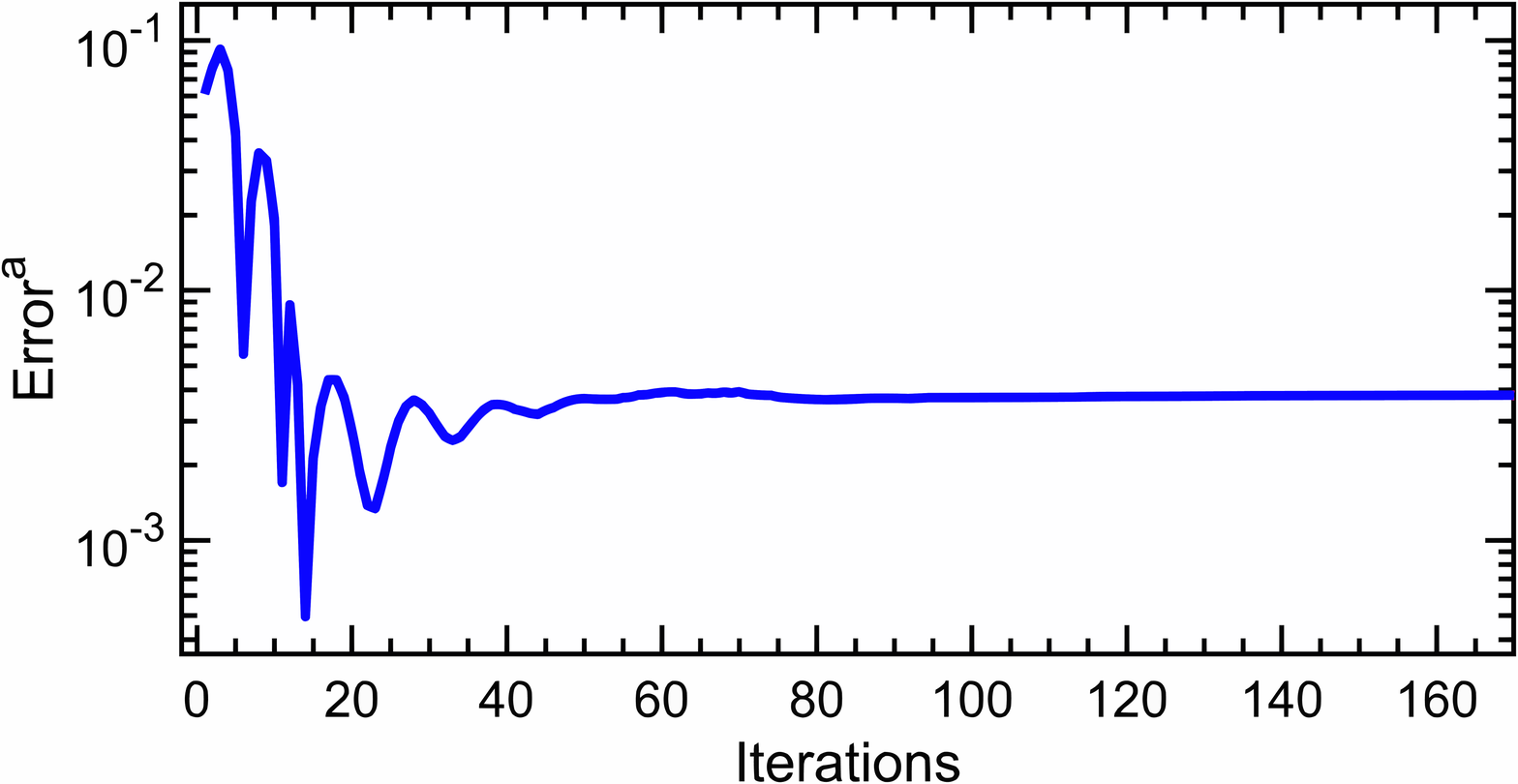}
		\caption{}
		\label{pic_conv}
	\end{subfigure}
	
	\begin{subfigure}{0.8\linewidth}
		\centering
		\includegraphics[width=\linewidth]{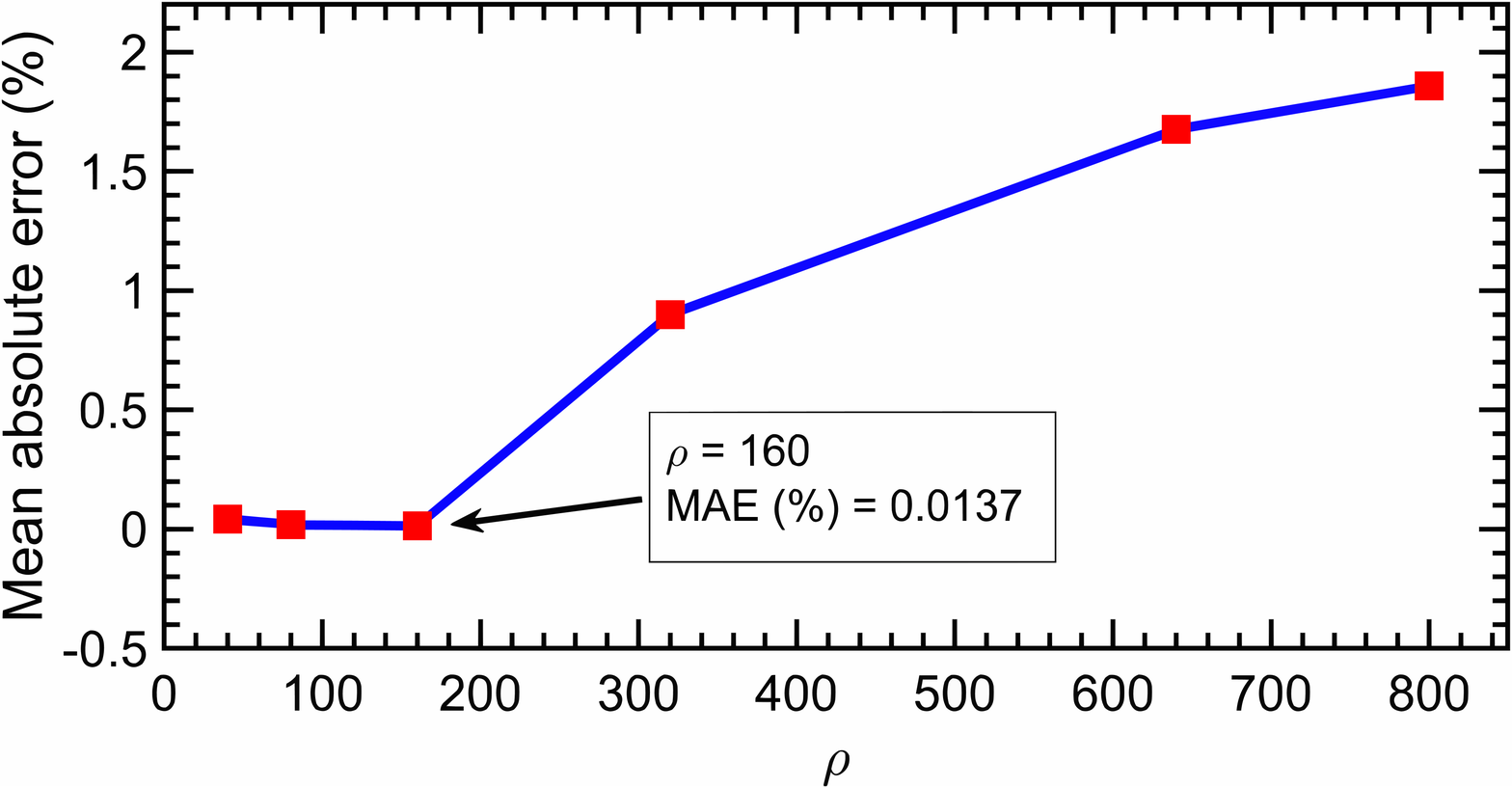}
		\caption{}
		\label{fig:mae_pic}
	\end{subfigure}
	\caption{(a) Convergence of shared variable $y_i^t$ (when $\rho= 160$ and $\epsilon=10^{-4}$), (b) convergence of error in optimal costs between centralized and distributed solutions (Error$^a$), (c) Error$^b$ with varying $\rho$ (when $\epsilon= 10^{-4}$).}
\end{figure}

	Fig. \ref{pic_con} depicts the convergence of shared variables $\mathbf{y}^t_i$ for each agent to the average $\mathbf{y}^{avg}$ value and shows that consensus among all agents is achieved for a given tolerance (as given in line 2 of Algorithm \ref{algo}). In Fig. \ref{pic_conv}, we also show the convergence of the error in optimal cost between centralized and distributed solution which is defined as
	\begin{align}
		\text{Error}^a = \frac{\mid C(\mathbf{x}^\star) - \{\hat{c}_0 + \sum_{i=1}^{N_{m}} \hat{c}_{i}\} \mid}{\mid C(\mathbf{x}^\star)\mid},
	\end{align}
	where $C(\mathbf{x}^\star)$ is the optimal cost \eqref{eq:total_cost} under centralized solution, $\hat{c}_0 :=  \sum^{N_p}_{k=1} \! \!  \big[c^{\top}_{0,k} \mathbf{z}_{0,k} + \bar{c}^{\top}_{0,k} \mathbf{y}_{0,k} \big] $ is defined from equation \eqref{sub_1} and $\hat{c}_i := \sum^{N_p}_{k=1} \big[c^{\top}_i \mathbf{z}_{i,k} + \bar{c}^{\top}_{i,k} \mathbf{y}_{i,k} \big] $ is defined from equation \eqref{sub_mg_1}. Error$^a$ is plotted in log scale and is found to be converging quickly. We also compute the error between the shared variables $y_i$ obtained from the solution of  the distributed algorithm and the centralized solution (as given in \eqref{cent_prob}).  This mean absolute error is defined as 
	\begin{align}
		\text{Error}^b = \left[ \frac{1}{(N_{m}+1).N_{sh}}\right]  \sum_{i=1}^{N_{m}+1} \sum_{j=1}^{N_{sh}}  \frac{\mid y^{cent}_j - y_{i,j} \mid}{\mid y^{cent}_j \mid},
	\end{align}
	where $y^{cent}_j$ is the $j$-th element of the shared variable vector $y^{cent}$ which can be obtained from \eqref{cent_prob},  $y_{i,j}$ is the $j$-th element of $y_i$ vector that belongs to $i$-th agent and $N_{sh}$ is the total number of shared variables. The Error$^b$ is shown in Fig. \ref{fig:mae_pic} for different values of hyper-parameter $\rho$. The ADMM scheme terminates when the consensus error becomes smaller than tolerance $\epsilon = 10^{-4}$ for each agent as given in line 2 of Algorithm \ref{algo}. The smallest deviation between the centralized and distributed solutions (i.e. Error$^b$) is obtained when $\rho = 160$ with the mean absolute error being $0.0137 \: \%$. 
	
	The per iteration  computation time for ADN and MG local optimization problems as well as the number of iterations required for convergence of ADMM are shown in Table \ref{table_dist} for different values of $\rho$ and error tolerance $\epsilon$. Please note that, computational time under the heading `MG' in Table \ref{table_dist} refers to the mean computational time per iteration required by each MG. It is observed that, computational time and number of iterations are smaller for higher values of $\rho$, i.e., the agents quickly reach consensus on the shared variables. Although Error$^a$ roughly remains same,  increase in $\rho$ leads to a larger mean absolute error (Error$^b$) between distributed and centralized solutions (as shown in Fig. \ref{fig:mae_pic}). Therefore, users need to choose the value of $\rho$ carefully by evaluating the trade-off between faster convergence and greater accuracy. Table \ref{table_dist} provides much valuable insights to this end. 
 
    Similarly, smaller $\epsilon$ led to a larger number of iterations before convergence is achieved. For the problem considered in this work, value of $\rho$ between $100$ and $200$ and $\epsilon = 10^{-4}$ led to mean absolute error (Error$^b$) being less than $2\%$. Finally, we clarify that the results shown in Table \ref{table_dist} are to illustrate the relative impact of different parameters such as $\epsilon$ and $\rho$ on the number of iterations and computation time. The absolute value of computation time would depend on the choice of processor, choice of MILP algorithm, implementation framework, among others; specifically, a native C/C++ implementation would lead to a smaller computation time. 
	
	\begin{table}[]
		\centering
		\caption{Analysis of computational time for agents. }
		\addtolength{\tabcolsep}{-3pt}
		\begin{tabular}{@{}|c|c|c|c|c|c|@{}}
			\hline
			\multirow{3}{*}{Tolerance ($\epsilon$)} & \multirow{3}{*}{Value} & \multirow{3}{*}{Iterations} & \multicolumn{2}{c|}{\multirow{2}{*}{Computation time}} & \multirow{3}{*}{Error$^a$}\\[3pt]
			&	&	& \multicolumn{2}{c|}{per iteration (s)}& \\
			\cline{4-5}
			&    of $\rho$              &                   & ADN              & MG   &     ($\%$)        \\
			\hline
			\multirow{4}{*}{$10^{-2}$}   & 60              & 122               & 6.48           & 0.63  & 0.2268        \\
			& 140             & 114               & 5.6            & 0.63  & 0.2236        \\
			& 300             & 109               & 5.2            & 0.64  &  0.2255        \\
			& 500             & 102               & 4.85            & 0.64  & 0.1004         \\
			\hline
			\multirow{4}{*}{$10^{-3}$}   & 60              & 151               & 5.89            & 0.64  & 0.2267      \\
			& 140             & 137               & 5.21            & 0.64   & 0.2238         \\
			& 300             & 132               & 4.97            & 0.63  &   0.2255      \\
			& 500             & 121               & 4.78              & 0.63  &   0.1777      \\
			\hline
			\multirow{4}{*}{$10^{-4}$}   & 60              & 174              & 5.63            & 0.64  &  0.2239       \\
			& 140             & 170               & 5.11            & 0.64  &   0.2239      \\
			& 300             & 163               & 4.81            & 0.63  &   0.2257     \\
			& 500             & 154               & 4.51             & 0.63  &     0.2254    \\
			
			\hline
			
		\end{tabular}
		\label{table_dist}
	\end{table}

\subsection{Scalability of the Proposed Approach}
To investigate the scalability of our formulation, we have carried out additional simulations on the IEEE $69$-bus \cite{matpower} DN, $136$-bus DN, and European low voltage test feeder 906-bus \cite{feeders2011ieee} DN. In addition, all the distribution systems are further populated with an increasing number of MGs. Table \ref{table_scalability} shows the number of iterations and per iteration computation time required by the agents to achieve convergence. In addition, the Error$^b$ between the centralized and distributed solutions are also shown. To accelerate the convergence process with a larger DN, we have chosen penalty factor $\rho = 160$ upto the point when tolerance ($\epsilon$) reaches $10^{-2}$ and thereafter, $\rho$ is increased to $1000$ until tolerance ($\epsilon$) reaches $10^{-4}$. 

The table shows that the number of iterations increases as number of MGs grows, primarily due to the increase in the dimension of the shared variables over which agents must achieve consensus. Despite this, the per iteration computation time of the MGs remains relatively same, as the size of the sub-problem \eqref{subproblem_mg} remains unchanged. However, a notable increase in per-iteration computation time for ADN is observed when transitioning to larger networks (where $N_p=10$), as the size of the sub-problem \eqref{eq:ADN_ADMM} significantly expands due to increase in power flow constraints and line flow limits. The increase is polynomial in nature since there is no increase in the dimension of the integer variables with increase in the network size. Additionally, while the sub-problem \eqref{subproblem_mg} is a Quadratic Programming (QP) instance, the sub-problem \eqref{eq:ADN_ADMM} of ADN belongs to the MIQP category. As a consequence, the solver MOSEK inherently requires more time to solve it. For the 69-bus case, the per iteration computation time of ADN sub-problem is relatively higher (even close to 136-bus case) because, the feasibility set of \eqref{eq:ADN_ADMM} resulted activation of majority number of integer variables (out of $3N_p$) which in turn necessitates MOSEK to solve more number of problems out of total $2^{3N_p}$ problems.

\begin{figure}
        \centering
        \includegraphics[width=0.9\linewidth]{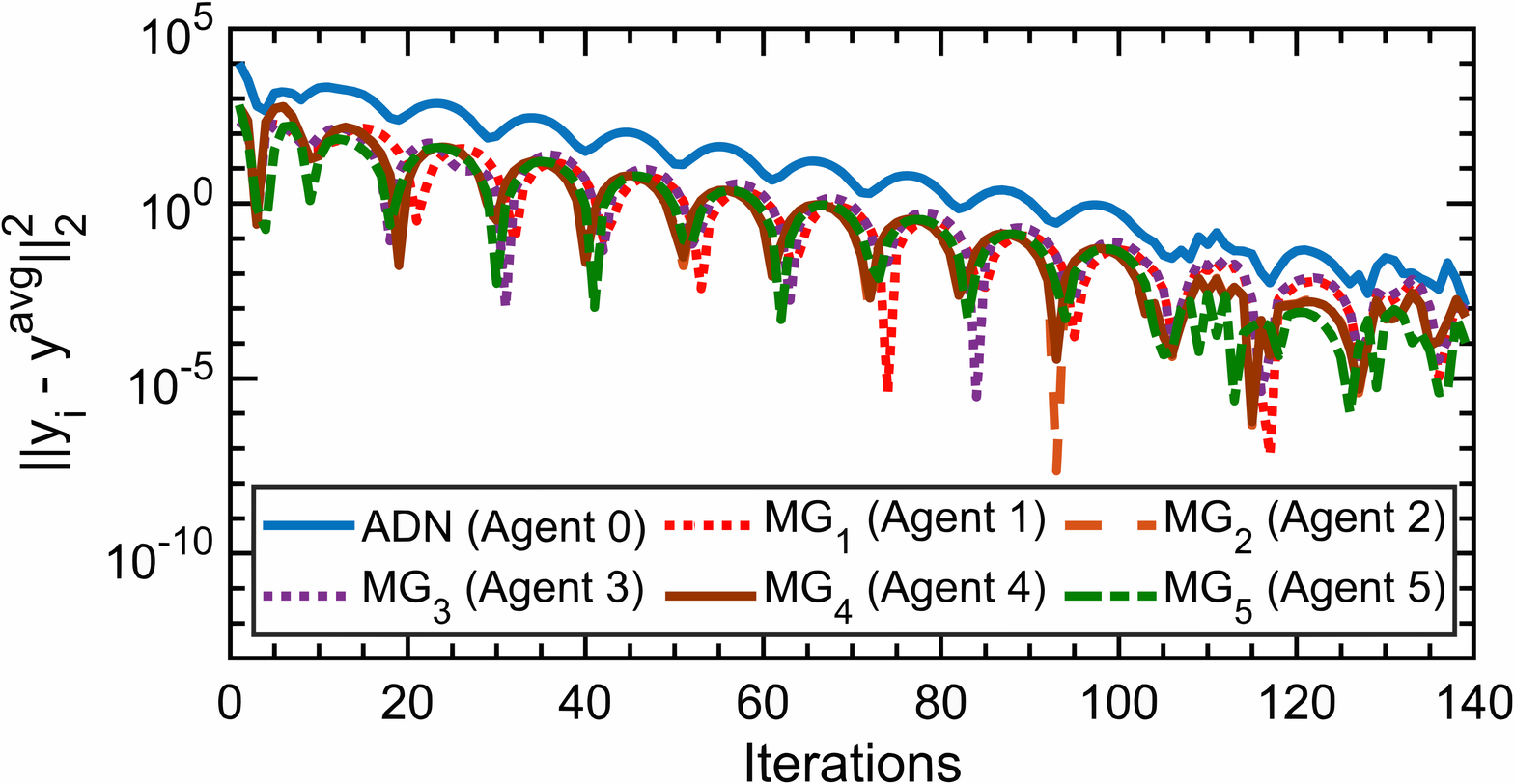}
        \caption{Convergence of shared variable $y^t_i$ (when $\epsilon=10^{-4}$) for European low voltage test feeder network of $906$ buses.}
        \label{conv_906bus}
    \end{figure}

The total computation time for the European low voltage test feeder ($906$-bus) Distribution Network (DN) was found to be significantly larger. In order to alleviate the computational burden, we reduced $N_p$ from $10$ to $2$ to reduce the dimension of integer variables in the optimization problem. This resulted in a significant acceleration of per-iteration computation time for ADN, which is evident from the table. Fig. \ref{conv_906bus} shows the convergence behavior of the distributed framework for this $906$-bus network with $5$ microgrids.

	\begin{table}[]
		\centering
		\caption{Computation time for different network sizes}
		\addtolength{\tabcolsep}{-3pt}
		\begin{tabular}{@{}|c|c|c|c|c|c|@{}}
			\hline
			\multirow{3}{*}{Test system} & \multirow{3}{*}{No. of} & \multirow{3}{*}{Iterations} & \multicolumn{2}{c|}{\multirow{2}{*}{Computation time}} & \multirow{3}{*}{Error$^b$}  \\ [3pt]
			&        &      &  \multicolumn{2}{c|} {per iteration (s)} &  	\\
			\cline{4-5}
			&	MGs 	  &         & ADN              & MG  &      ($\%$)         \\
			\hline
			\multirow{3}{*}{IEEE 33-bus ($N_p = 10$)} &	5 & 156        & 4.16       & 0.63         & 0.96           \\
			&	7 & 169        & 4.65        & 0.82       & 0.77           \\
			&	10 & 197       & 4.62      & 0.99        & 1.22          \\
			\hline						
			\multirow{3}{*}{IEEE 69-bus ($N_p = 10$)} &	5 & 135       & 17.77       & 0.61        & 1.11           \\
			&	7 & 181       & 15.49       & 0.84        & 1.21           \\
			&	10 & 246			&	13.77		&	1.13	&	1.76			\\
			\hline
   			\multirow{3}{*}{136-bus DN ($N_p = 10$)} &	5 & 126       & 14.41      & 0.927        & 0.007           \\
			&	7 & 175       & 13.414       & 1.09        & 0.01           \\
			&	10 & 196			&	13.79		&	1.43	&	0.014			\\
			\hline
			\multirow{3}{*}{ELV 906-bus ($N_p = 2$)} &	5 & 140       & 25.106      & 0.205        & 0.0145           \\
			&	10 & 180       & 26.681       & 0.1        & 0.824           \\
			&	20 & 259			&	25.75		&	0.173	&	1.39			\\
			\hline
		\end{tabular}
		\label{table_scalability}
	\end{table}

\section{Summary and Conclusion}
	In this paper, an ADMM based fully distributed optimization framework is proposed to coordinate multi-MGs in an ADN and provide AS to the TN while satisfying the operational constraints of the network.  We presented detailed numerical results on a benchmark distribution network and showed that participating in the passive voltage support scheme does not lead to any significant increase in the operational cost of the ADN.  The effect of ADMM hyper-parameters on convergence time and accuracy are thoroughly investigated. The computation time for the ADN and MG sub-problems, and the impact of network size and prediction horizon, are also reported demonstrating the scalability of our approach.
 
	While passive voltage support based ancillary service, as recommended by the Belgian TSO, is investigated in this work, the proposed formulation is easily amenable to other types of ancillary services (for instance, Swiss TSO model \cite{stavros_as}) as well. Similarly, active voltage support schemes is also realizable in the proposed framework, and will be examined in a follow-up work. In addition, designing distributed algorithms that incorporate the effect of uncertainty associated with renewable energy generation and enable peer-to-peer trading among microgrids while providing ancillary services remain as promising directions for future research.  Finally, developing distributed variants of algorithms such as particle swarm optimization for constrained multi-agent optimization problems that arise in power systems applications should be investigated in future work.

	\bibliographystyle{IEEEtran}
	\bibliography{main.bib}
	
\end{document}